\documentclass[paper,notoc,12pt]{JHEP3}

\usepackage{amsmath}
\usepackage{graphicx}
\usepackage[sort&compress,numbers]{natbib}

\usepackage{psfrag}
\usepackage{epsf}

\def\be{\begin{equation}}
\def\ee{\end{equation}}
\def\bea{\begin{eqnarray}}
\def\eea{\end{eqnarray}}

\def\Neqfour{{\cal N}=4}
\def\Neqone{{\cal N}=1}

\def\spa#1.#2{\left\langle#1\,#2\right\rangle}
\def\spb#1.#2{\left[#1\,#2\right]}
\def\lor#1.#2{\left(#1\,#2\right)}
\def\sand#1.#2.#3{%
\left\langle\smash{#1}{\vphantom1}^{-}\right|{#2}%
\left|\smash{#3}{\vphantom1}^{-}\right\rangle}
\def\sandp#1.#2.#3{%
\left\langle\smash{#1}{\vphantom1}^{-}\right|{#2}%
\left|\smash{#3}{\vphantom1}^{+}\right\rangle}
\def\sandpp#1.#2.#3{%
\left\langle\smash{#1}{\vphantom1}^{+}\right|{#2}%
\left|\smash{#3}{\vphantom1}^{+}\right\rangle}
\def\sandpm#1.#2.#3{%
\left\langle\smash{#1}{\vphantom1}^{+}\right|{#2}%
\left|\smash{#3}{\vphantom1}^{-}\right\rangle}
\def\sandmp#1.#2.#3{%
\left\langle\smash{#1}{\vphantom1}^{-}\right|{#2}%
\left|\smash{#3}{\vphantom1}^{+}\right\rangle}
\def\sandmm#1.#2.#3{%
\left\langle\smash{#1}{\vphantom1}^{-}\right|{#2}%
\left|\smash{#3}{\vphantom1}^{-}\right\rangle}
\def\spab#1.#2.#3{\sandmm#1.#2.#3}
\def\spbb#1.#2.#3.#4{\sandpm#1.{#2#3}.#4}
\newbox\charbox
\newbox\slabox
\def\s#1{{      
        \setbox\charbox=\hbox{$#1$}
        \setbox\slabox=\hbox{$/$}
        \dimen\charbox=\ht\slabox
        \advance\dimen\charbox by -\dp\slabox
        \advance\dimen\charbox by -\ht\charbox
        \advance\dimen\charbox by \dp\charbox
        \divide\dimen\charbox by 2
        \raise-\dimen\charbox\hbox to \wd\charbox{\hss/\hss}
        \llap{$#1$}
}}

\newcommand{\DD}{\mathrm{DD}}

\def\beqa{\begin{eqnarray}}
\def\eeqa{\end{eqnarray}}
\def\beq{\begin{equation}}
\def\eeq{\end{equation}}

\def    \br#1#2          {\mbox{$\langle #1 \, #2 \rangle$}}
\def    \sq#1#2          {\mbox{$\left[  #1 \, #2 \right]$}}
\def    \sap#1#2#3       {\mbox{$\langle #1 | #2 |#3  \rangle$}}
\def    \t#1#2#3         {\mbox{$s_{#1 #2 #3}$}}
\def    \s#1#2           {\mbox{$s_{#1 #2}$}}
\def    \sapp#1#2#3#4    {\mbox{$\langle #1 | (#2+#3) |#4  \rangle$}}

\def    \br(#1,#2)          {\mbox{$\langle #1 \, #2 \rangle$}}
\def    \sq(#1,#2)          {\mbox{$\left[  #1 \, #2 \right]$}}
\def    \t(#1,#2,#3)        {\mbox{$s_{#1 #2 #3} $}}
\def    \s(#1,#2)           {\mbox{$s_{#1 #2}$ }}
\def    \Pperp(#1)          {\mbox{$k_{#1 \perp}$ }}
\def    \PperpStar(#1)      {\mbox{$k_{#1 \perp}^*$ }}
\def    \Pperpt(#1)          {\mbox{$|k_{#1 \perp}|^2$ }}
\def    \x(#1)              {\mbox{$x_{#1}$ }}

\newcommand{\ssplit}[2]{\mathrm{split}(#1\to P^{#2})}
\newcommand{\Split}{\mathrm{Split}^{(n)}}

\def\Del#1.#2.#3{\Delta_{(1)}(#1,#2;#3)}
\def\Delq#1.#2{\Delta_{(1)}(#1;#2)}

\def\Dell#1.#2.#3.#4{\Delta_{(2)}({#1,#2};{#3,#4})}
\def\Dellq#1.#2{\Delta_{(2)}(q_{#1},q_{#2})}

\newcommand{\aab}[2]{\ensuremath{\langle#1 \,#2\rangle}}
\newcommand{\ssb}[2]{\ensuremath{[#1 \,#2]}}
\newcommand{\asb}[3]{\ensuremath{\langle#1 |#2 |#3]}}

\allowdisplaybreaks[1]

\author{
T.~G. Birthwright, E.~W.~N.~Glover, V.~V.~Khoze and P.~Marquard.\\
Department of Physics,
University of Durham,
Durham DH1 3LE,
U.K.\\
E-mail:  \email{T.G.Birthwright@durham.ac.uk, E.W.N.Glover@durham.ac.uk,
Valya.Khoze@durham.ac.uk, Peter.Marquard@durham.ac.uk}
}

\title{Multi-Gluon Collinear Limits from MHV diagrams}

\preprint{DCPT/05/04, IPPP/05/02, hep-ph/0503063}

\abstract{
We consider the multi-collinear limit of multi-gluon QCD
  amplitudes at tree level. We use the MHV rules for constructing
  colour ordered tree amplitudes and the general collinear
  factorization
  formula to derive timelike splitting functions
  that are valid for specific numbers of negative helicity
  gluons and an arbitrary number of positive helicity gluons (or vice versa).
  As an example
  we present new results describing
  the collinear limits of up to six gluons.  }

\begin{document}

\section{Introduction}
\label{sec:intro}

The interpretation of $\Neqfour$ supersymmetric Yang-Mills theory and QCD as a
topological string propagating in twistor space~\cite{Witten1}, has inspired a
new and powerful framework for computing tree-level and one-loop
scattering amplitudes in  Yang-Mills gauge theory.
Notably, two distinct formalisms have been developed for calculations of
scattering amplitudes in gauge theory -- the `MHV rules' of Cachazo, Svr\v{c}ek and Witten
(CSW) \cite{CSW1}, and the `BCF recursion relations' of Britto, Cachazo, Feng and Witten
\cite{BCF4,BCFW}.

In this paper, we wish to exploit these formalisms to examine the singularity
structure of tree-level amplitudes
when many gluons are simultaneously collinear. Understanding the
infrared singular behaviour of multi-parton amplitudes is a prerequisite for
computing infrared-finite cross sections at fixed order in perturbation
theory.  In general, when one or more final state particles are either soft or
collinear, the amplitudes factorise. The first factor in this product is
an amplitude depending
on the remaining hard partons in the process (including any hard partons
constructed from an ensemble of unresolved partons). The second factor
contains
all of the singularities due to the unresolved particles.   One of the best
known examples of this type of factorisation is the limit of tree amplitudes
when two particles are collinear.  This factorisation is universal and can be
generalised to any number of loops~\cite{Kosower:allorderfact}.

Both frameworks, the MHV rules and the BCF recursion relations, are remarkably powerful
in deriving analytic expressions for massless multi-particle tree-level amplitudes.
At the same time, for the specific purpose of deriving general multi-collinear limits,
we find the MHV rules approach to be particularly convenient.

A useful feature of the MHV rules  is that it is not required to set
reference spinors $\eta_\alpha$ and $\eta_{\dot\alpha}$
to specific values dictated by kinematics or other reasons.
In this way, on-shell (gauge-invariant) amplitudes are derived for
arbitrary $\eta$'s, i.e. without fixing the gauge.
By starting from the appropriate colour ordered
amplitude and taking the collinear limit,  the full amplitude factorises  into an
MHV vertex multiplied by a multi-collinear splitting function that depends on
the helicities of the collinear gluons.
Because the MHV vertex is a single factor, the collinear
splitting functions have a similar structure to MHV amplitudes.  Furthermore,
the gauge or $\eta$-dependence of the splitting function drops out.

One of the main points of our approach is that, in order to derive all required splitting
functions we do not need to know the full amplitude. Out of the full set of
MHV-diagrams contributing to the full amplitude, only a subset will contribute
to the multi-collinear limit. This subset includes only those MHV-diagrams which contain
an internal propagator which goes on-shell in the multi-collinear limit.
In other words, the IR singularities in the MHV approach arise entirely from internal
propagators going on-shell. This observation is specific to the MHV rules method
and does not apply to the BCF recursive approach. We will see in Section {\bf \ref{sec:4.2.3}}
that in the BCF picture collinear splitting functions generically
receive contributions from the full set of allowed BCF diagrams\footnote{This
is because the required IR poles in the BCF approach arise not only from propagators
going on-shell, but also from the constituent BCF vertices.}.
In view of this, we will employ the MHV rules of \cite{CSW1} for setting up the
formalism and for derivations of general multi-collinear amplitudes.
At the same time, various specific examples of multi-collinear splitting amplitudes
derived in this paper will also be checked in Section {\bf \ref{sec:4.2}} using
the BCF recursion relations \cite{BCF4}.

\medskip

The basic building blocks
of the MHV rules approach~\cite{CSW1}
are the colour-ordered
$n$-point vertices  which are connected by scalar propagators. These MHV
vertices are off-shell continuations of the maximally helicity-violating (MHV)
$n$-gluon scattering amplitudes  of Parke and Taylor~\cite{ParkeTaylor,BG}.   They
contain precisely two negative helicity gluons.
Written in terms of spinor inner
products~\cite{SpinorHelicity}, they are composed entirely of the holomorphic
products $\spa{i}.{j}$ of the right-handed (undotted) spinors,
rather than their anti-holomorphic partners $\spb{i}.{j}$,
\be
 A_n(1^+,\ldots,p^-,\ldots,q^-,\ldots,n^+) =
\frac{\spa{p}.{q}^4}{ \spa1.2 \spa2.3 \cdots \spa{n-1,}.{n} \spa{n}.{1} },
\label{MHV}
\ee
where we introduce the common notation
$\spa{p_i}.{p_j}=\spa{i}.{j}$ and $\spb{p_i}.{p_j}=\spb{i}.{j}$.
By connecting MHV vertices,  amplitudes involving more
negative helicity gluons can be built up.

The MHV rules for gluons~\cite{CSW1} have been extended to amplitudes with
fermions~\cite{GK}.  New compact results for tree-level
gauge-theory results for non-MHV amplitudes  involving arbitrary numbers of
gluons~\cite{Zhu,KosowerNMHV,BBK}, and fermions~\cite{GK,GGK,Wu1,Wu2} have been
derived.  They have been applied to processes involving external Higgs
bosons~\cite{DGK,BGK} and electroweak bosons~\cite{BFKM}.
MHV rules for tree amplitudes have further been recast in the
form of recursive relations
~\cite{BBK,BFKM,BGK}
which facilitate calculations of higher order non-MHV amplitudes in terms of the
known lower-order results.
In many cases new classes of tree amplitudes were derived, and
in all cases, numerical agreement with
previously known amplitudes has been found.

MHV rules have also been shown to work at the loop-level for
supersymmetric theories.
Building on the earlier work of Bern et al~\cite{BDDK1,BDDK2},
 there has been enormous progress in computing cut-constructible multi-leg
loop amplitudes in $\Neqfour$~\cite{CSW2,BST,CSW3,BBKR,Cachazo,
BCF1,BDDK7,BCF2,BCF3,BDKNMHV}
and $\Neqone$~\cite{QuigleyRozali,BBST1,BBDD,BBDP1}
supersymmetric gauge theories.  Encouraging
progress has also been made for non-supersymmetric loop
amplitudes~\cite{BBST2, BBDP2, BDKrec}.

Remarkably, the expressions obtained for the infrared singular parts of
$\Neqfour$ one-loop amplitudes (which are known to be  proportional to
tree-level results) were found to produce even more compact expressions for
gluonic tree amplitudes~\cite{BDKNMHV,RSV}. This observation led to the BCF
recursion relations~\cite{BCF4,BCFW} discussed earlier as well as extremely
compact six-parton amplitudes~\cite{LW1,LW2} and expressions  for MHV and NMHV
graviton amplitudes~\cite{BBST3,CS}.

\medskip

The factorisation properties of amplitudes in the infrared
play several roles in developing higher order
perturbative predictions for observable quantities. First, a detailed
knowledge of the structure of unresolved emission enables phase space
integrations to be organised such that the infrared singularities due to soft
or collinear emission can be analytically
extracted~\cite{Giele:1992vf,Frixione:1996ms,Catani:1997vz}.    Second, they
enable large logarithmic corrections to be identified and resummed.
Third, the collinear limit plays a crucial role in the unitarity-based method
for loop calculations~\cite{BDDK1,BDDK2,Bern:1996db,Bern:1996je}.

In general, to compute a cross section at N$^n$LO, one requires detailed
knowledge of the infrared factorisation functions describing the unresolved
configurations for $n$-particles at tree-level, $(n-1)$-particles at one-loop
etc. The universal behaviour in the double collinear limit is well known at
tree-level (see for example Refs.~\cite{Altarelli:1977zs,Bassetto:1983ik}),
one-loop~\cite{Bern:1995ix,BDDK1,Bern:split1gluon,
Kosower:split1,Bern:split1QCD,Catani:2000pi} and at
two-loops~\cite{Bern:2lsplit,Badger:2lsplit}. Similarly, the triple collinear
limit has been studied at
tree-level~\cite{Gehrmann-DeRidder:dblunres,Campbell:dblunres,Catani:NNLOcollfact,Catani:IRtreeNNLO} and,
in the case of distinct quarks, at one-loop~\cite{Catani:2003}. Finally, the
tree-level quadruple gluon collinear limit is derived in Ref.~\cite{delduca}.

\medskip

Our paper is organised as follows.   In Section~{\bf \ref{sec:mhv}}, we briefly review
the spinor helicity and colour ordered formalism that underpins the MHV
rules.  Section~{\bf \ref{sec:limit}} describes the procedure for taking the collinear
limit and deriving the splitting functions.  We write down a general collinear
factorization formula, which is valid for specific numbers of negative
helicity gluons and an arbitrary number of positive helicity gluons and
demonstrate that the gauge dependence explicitly cancels.    We find  it
useful to classify our results according to the difference between the number
of negative helicity gluons before taking the collinear limit, and the number
after. We call this difference $\Delta M$. We provide formulae describing an
arbitrary number of gluons  for $\Delta M \leq 2$ in Section~{\bf \ref{sec:gen}}.
Specific explicit results for the collinear limits of up to six gluons are given in
Sec.~{\bf \ref{sec:expl}}.   We have numerically checked that our results agree with
the results available in the literature for three and four collinear
gluons~\cite{delduca}. Our findings are summarized in Sec.~{\bf \ref{sec:concl}}.

\section{Colour ordered amplitudes in the spinor helicity formalism}
\label{sec:mhv}

Tree-level multi-gluon amplitudes can be decomposed into
colour-ordered partial amplitudes as
\begin{equation}
{\cal A}_n(\{p_i,\lambda_i,a_i\}) =
i g^{n-2}
\sum_{\sigma \in S_n/Z_n} {\rm Tr}(T^{a_{\sigma(1)}}\cdots T^{a_{\sigma(n)}})\,
A_n(\sigma(1^{\lambda_1},\ldots,n^{\lambda_n}))\,.
\label{TreeColourDecomposition}
\end{equation}
Here $S_n/Z_n$ is the group of non-cyclic permutations on $n$
symbols, and $j^{\lambda_j}$ labels the momentum $p_j$ and helicity
$\lambda_j$ of the $j^{\rm th}$ gluon, which carries the adjoint
representation index $a_i$.  The $T^{a_i}$ are fundamental
representation SU$(N_c)$ colour matrices, normalized so that
${\rm Tr}(T^a T^b) = \delta^{ab}$.  The strong coupling constant is
$\alpha_s=g^2/(4\pi)$.
The MHV rules method of Ref.~\cite{CSW1} is used to evaluate only the
purely kinematic amplitudes $A_n.$
Full amplitudes are then determined uniquely from the kinematic part $A_n$,
and the known expressions for the colour traces.

In the spinor helicity formalism
\cite{SpinorHelicity,ParkeTaylor,BG} an on-shell momentum
of a massless particle, $p_\mu p^\mu=0,$ is represented as
\be
p_{a \dot a} \equiv \ p_\mu \sigma^\mu_{a \dot a}
=\ \lambda_a\tilde\lambda_{\dot a} \ ,
\ee
where $\lambda_a$ and $\tilde\lambda_{\dot a}$
are two commuting
spinors of positive and negative chirality.
Spinor inner products are defined
by\footnote{Our conventions for spinor helicities follow
\cite{Witten1,CSW1}, except that $[ij] = - [ij]_{CSW}$
as in ref.~\cite{LDTASI}.}
\be
\langle \lambda,\lambda'\rangle = \ \epsilon_{ab}\lambda^a\lambda'{}^b
 \ , \qquad
[\tilde\lambda,\tilde\lambda'] =\ -\epsilon_{\dot a\dot b}
\tilde\lambda^{\dot a}\tilde\lambda'{}^{\dot b} \ ,
\ee
and a scalar product of two null vectors,
$p_{a\dot a}=\lambda_a \tilde\lambda_{\dot a}$ and
$q_{a\dot a}=\lambda'_a\tilde\lambda'_{\dot a}$, becomes
\be \label{scprod}
p_\mu q^\mu =\ -\frac{1}{2}
\langle\lambda,\lambda'\rangle[\tilde\lambda,\tilde\lambda'] \ .
\ee

The MHV rules of Ref.~\cite{CSW1}
were developed for calculating
purely gluonic amplitudes at tree level. In this approach
all non-MHV $n$-gluon
amplitudes (including $\overline{\rm MHV}$) are expressed
as sums of tree diagrams in an effective scalar perturbation theory.
The vertices in this theory are the MHV amplitudes of Eq.~(\ref{MHV})
continued off-shell as described below, and connected by scalar
propagators $1/q^2$.

When one leg of an MHV vertex is connected by a propagator
to a leg of another MHV vertex, both legs become internal
to the diagram and have to be continued off-shell. Off-shell continuation
is defined as follows \cite{CSW1}:
we pick an arbitrary reference spinor $\eta^{\dot a}$ and define
$\lambda_a$ for any internal line carrying momentum $q_{a\dot a}$
by
\be \label{ofsh}
\lambda_a=q_{a\dot a}\eta^{\dot a}\ .
\ee
External lines in a diagram remain on-shell, and for them
$\lambda$ is defined in the usual way.
For the off-shell lines,
the same reference spinor $\eta$ is used  in all diagrams
contributing to a given amplitude.

\section{The multiple collinear limit}
\label{sec:limit}

To find the splitting functions we work with the colour stripped  amplitudes
$A_n$. For these colour ordered amplitudes, it is known  that when the
collinear particles are not adjacent there is no collinear
divergence~\cite{delduca}. Therefore, without loss of generality,  we can take
particles $1 \dots n$ collinear.

The multiple collinear limit is approached when the momenta $p_1,
\dots, p_n$ become parallel.  This implies that all the
particle subenergies $s_{ij}=(p_i+p_j)^2$, with $i,j=1,\dots,n$, are
simultaneously small. We thus introduce a pair of
light-like momenta $P^\nu$
and $\xi^\nu$ ($P^2=0, \xi^2=0$), and we write
\begin{equation}
(p_1 + \dots + p_n)^\nu = P^\nu
+ \frac{s_{1,n} \; \xi^\nu}{2 \, \xi \cdot P} \;, \quad
s_{i,j} = (p_i + \dots + p_j)^2 \;,
\end{equation}
where $s_{1,n}$ is the total invariant mass of the system of collinear
partons. In the collinear limit, the vector $P^\nu$
denotes the collinear direction, and the individual collinear momenta are
$p_i^\nu \to z_i P^\nu$. Here the longitudinal-momentum
fractions $z_i$ are given by
\begin{equation}
z_i = \frac{\xi \cdot p_i}{\xi \cdot P}
\end{equation}
and fulfil the constraint $\sum_{i=1}^m z_i =1$.
To be definite, in the rest of the paper we work
in the time-like region so that
($s_{ij} > 0, \;  1>z_i > 0$).

\begin{figure}[htbp]
  \centering
  \psfrag{-P-l}{$-P^{-\lambda}$}
  \psfrag{Pl}{$P^{\lambda}$}
  \psfrag{pn+1}{$(n+1)$}
  \psfrag{pn}{$n$}
  \psfrag{pN}{$N$}
  \psfrag{p1}{$1$}
  \includegraphics[width=12cm]{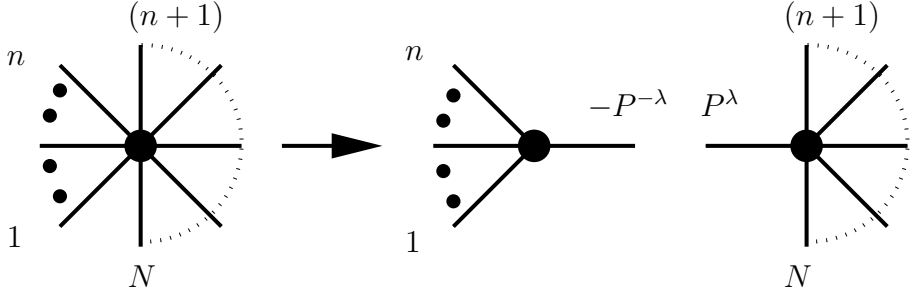}
  \caption{Factorisation of an $N$-point colour ordered amplitude with gluons $p_1,\ldots,p_n$ collinear
  into splitting function for $P \to 1, \ldots, n$
  multiplied by an $(N-n+1)$-point amplitude.}
  \label{fig:limit}
\end{figure}
As illustrated in Fig.~\ref{fig:limit},
in the multi-collinear limit an $N$-gluon colour ordered tree amplitude factorises
and can be written as
\bea
  \label{eq:factorise}
  A_N(1^{\lambda_1},\ldots,N^{\lambda_N}) &\to&
  \ssplit{{1}^{\lambda_{1}},\ldots,n^{\lambda_n} }{\lambda}
  \times
  A_{N-n+1}((n+1)^{\lambda_{n+1}},\ldots ,N^{\lambda_N},P^\lambda).\nonumber \\
\eea
This labelling of the splitting amplitude
$\ssplit{1^{\lambda_1},\ldots,n^{\lambda_n}}{\lambda}$ differs from the
usual definition because we use the momentum and helicity that
participates in the resultant amplitude $P^\lambda$ rather than $-P^{-\lambda}$.
With this choice, it is easier to see how the helicity is conserved
in the splitting, i.e.   helicity $\lambda^1,\ldots,\lambda^n$ is replaced by
$\lambda$.

There are two different types of collinear limit~\cite{CSW1},
those that conserve the number of negative helicity gluons between the
initial state and the final collinear state,
and those that do
not.

Only the limits of the type
$\ssplit{1^+,\ldots,n^+}{+}$ and
$\ssplit{1^-,2^+,\ldots,n^+}{-}$ can contribute to the negative helicity
conserving case,  and these
collinear splitting functions
 are straightforward to derive directly from the simple MHV vertex.

All other
limits belong to the second class which do not
conserve the number of negative helicity gluons, and therefore we classify
our results according to the difference between the number of negative
helicity gluons before taking the collinear limit, and the number after,
$\Delta M$. We find that $\Delta M$ corresponds to the
order of MHV diagram needed to find a particular collinear limit, as follows,
\bea
\Delta M=0 \Rightarrow  \phantom {NNN}{\rm MHV}:
&&1^+,2^+,3^+, \ldots ,n^+  \rightarrow P^+ \nonumber\\
&&1^-,2^+,3^+, \ldots ,n^+  \rightarrow P^-\nonumber\\
    & &\nonumber\\
\Delta M=1 \Rightarrow \phantom{NN}{\rm NMHV}:
&&1^-,2^+,3^+, \ldots ,n^+  \rightarrow P^+ \nonumber\\
&&1^-,2^-,3^+, \ldots ,n^+  \rightarrow P^- \label{mmp} \nonumber\\
  & &\nonumber\\
\Delta M=2 \Rightarrow  \phantom{N}{\rm NNMHV}:
&&1^-,2^-,3^+, \ldots ,n^+  \rightarrow P^+\nonumber\\
&&1^-,2^-,3^-, \ldots ,n^+  \rightarrow P^-\nonumber\\
\eea
and so on for all $\Delta M>2$ cases.

The splitting functions are derived by examining the general form of MHV
diagrams, which consist of MHV vertices and scalar propagators.
The general form of the $n$-particle collinear splitting functions is given by
 \bea
 \label{dmzero}
 \Delta M =0:{\   } &\propto& \frac{1}{\spa{\ }.{\ }^{n-1}},\\
 \label{dmnzero}
 \Delta M \neq 0:{\   } &\propto& \frac{1}{\spb{\ }.{\ }^{v-1} \spa{\ }.{\
 }^{n-v}},
 \eea
such that $(v-1)+(n-v)=n-1$, where $v$ is the number of vertices, and thus
$v-1$ is the number of scalar propagators. From \eqref{dmnzero} it follows that
for an MHV-diagram to contribute to $ \Delta M \neq 0$ collinear limits, it
is required to contain anti-holomorphic spinor products $\spb{i}.{j}$ of
collinear momenta. However, because on-shell MHV vertices are 
entirely holomorphic,
 within the MHV rules there are only two potential sources of the
anti-holomorphic spinor products. One source is scalar propagators
$1/s_{ij}=1/\spa{i}.{j} \spb{j}.{i}$ which inter-connect MHV vertices. The
second source is the off-shell continuation of the corresponding connected
legs in the MHV vertices.  Each off-shell continued leg of momentum $P$ gives
rise to a factor $\langle i  P \rangle \propto \langle i | P | \eta]$ which
gives rise to anti-holomorphic factors of $[j \eta]$. When the reference
spinors $\eta_{\dot\alpha}$ are kept general, and specifically, not set to be
equal to one of the momenta in the collinear set, the
$\eta$-dependence must cancel and the off-shell continuation cannot give rise
to an overall factor of $\spb{i}.{j}$. 

Therefore, the only source of singular anti-holomorphic factors are 
MHV-diagrams that contain an internal propagator of momentum 
$q_{i+1,j}=p_{i+1}+\ldots + p_j$ which is a sum of external momenta from the
collinear set such that $q^2=s_{i+1,j} \rightarrow 0$. Hence, we conclude that
only a subset of MHV-diagrams contributes to multi-collinear limits of tree
amplitudes. The subset is determined by requiring that all $v-1$ internal
propagators are on-shell in the multi-collinear limit. This is a powerful
constraint on the types of the contributing diagrams and it simplifies the
calculation\footnote{Note that this selection rule would not apply to neither
gauge-fixed MHV rules (where $\eta$'s are fixed to be equal to kinematic
variables from the collinear set), nor to the BCF rules which mix holomorphic
MHV vertices with anti-holomorphic $\overline{\rm MHV}$ ones.}.

We exploit the universal nature of the splitting function by choosing to start
with an amplitude with $(n+3)$ external legs, i.e. setting $N=n+3$ in
Eq.~(\ref{eq:factorise}). The helicities of the gluons are adjusted so
that the remnant `hard'  four point MHV amplitude $A_4(P^\lambda, (n+1)^+,
(n+2)^{-\lambda},  (n+3)^-)$ is given by
\bea
\lefteqn{A_4( (n+1)^+, (n+2)^{-\lambda}, 
(n+3)^-,P^\lambda)=\hspace{3cm}\phantom{~}}\nonumber\\
\phantom{~}\hspace{1cm}&&\frac{\spa{n+3,}.{X}^4}
{\spa{P,}.{n+1}\spa{n+1,}.{n+2}\spa{n+2,}.{n+3}\spa{n+3,}.{P}}
\eea
with $X=P$ for $ \lambda=-$ and $X=n+2$ for $\lambda=+$.
  
To read off the collinear limits from the MHV rules, we use the limiting expressions
for the spinor products: $\spa{a}.{q}$, $\spa{b}.{q}$ and $\spa{b}.{a}$. Here $a$ is a
particle from the collinear set, 
$b$ is a particle which is not in the collinear set, and
$q$ is the sum of the collinear momenta from $i+1$ to $j$.
Hence, using
\be
  \spa{a}.{q}=\sum_{l=i+1}^{j}\spa{a}.{l} \spb{l}.{\eta} \ , \quad
  \spa{b}.{q}=\sum_{l=i+1}^{j}\spa{b}.{l} \spb{l}.{\eta} \ ,
\ee
and the expressions for spinors from the collinear set,
\be
|l\rangle = \sqrt{z_l} |P\rangle \ , \qquad
|l] = \sqrt{z_l} |P] \ , \qquad
|a\rangle = \sqrt{z_a} |P\rangle \ , \qquad
|a] = \sqrt{z_a} |P] \ ,
\ee
we have,
\bea
  \spa{a}.{q}&\rightarrow&\, \spb{P}.{\eta}\sum_{l=i+1}^{j}\spa{a}.{l}\sqrt{z_l} \, \equiv\,
  \spb{P}.{\eta}\Del{i}.{j}.{a} \label{aqqq}\\
  \spa{b}.{q}&\rightarrow&\, \spb{P}.{\eta}\spa{b}.{P}\sum_{l=i+1}^{j} z_l \label{Xq}\\
  \spa{b}.{a}&\rightarrow&\, \spa{b}.{P} \sqrt{z_a} .\label{Xa}
\eea
Here we introduced the definition
\begin{equation}
  \label{eq:7}
  \Del{i}.{j}.{a} = \sum_{l=i+1}^{j}\spa{a}.{l}\sqrt{z_l} .
\end{equation}

Equations~\eqref{aqqq} and \eqref{Xq}  contain a factor $\spb{P}.{\eta}$ which,
however, will always cancel
in expressions for relevant splitting functions.
As such we can read off the collinear limits of the amplitudes from
the MHV-rules expressions by replacing terms on the left hand
side of equations~\eqref{aqqq}, \eqref{Xq} and \eqref{Xa} with the expressions on the
right hand side of those equations, and further dropping the $\spb{P}.{\eta}$ factors.

Certain terms in the sums that arise in MHV rules need special attention.  These are the
 boundary terms involving either $\spa{0}.{1}$ or $\spa{n}.{n+1}$, 
 and for these we have,
\bea
 \frac{\spa{n}.{n+1}}{\Del{i}.{n}.{n+1}} &\rightarrow&\,
 -\frac{\sqrt{z_n}}{\sum_{l=i+1}^{n}z_l}, \label{jn}\\
 \frac{\spa{0}.{1}}{\Del{0}.{j}.{0}} &\rightarrow&\,
 \frac{\sqrt{z_1}}{\sum_{l=1}^{j}z_l}.\label{inp3} 
\eea
We now present our results for the splitting functions.

\section{Results}
\label{sec:results}
In this section we give the results for the multiple collinear limit
of gluons. First we give the general results for an arbitrary number
of gluons with $\Delta M \leq 2$. Afterwards we give explicit results
for up to four collinear gluons for all independent helicity combinations,
together with some specific examples for five and six collinear gluons.

\subsection{General results}
\label{sec:gen}

In this section we present the general results for the cases where
the number of gluons with negative helicity changes by at most $\Delta
M = 2$, and those related by parity where the number of gluons with
positive helicity changes at most by the same amount. With the help of
parity these general splitting amplitudes are sufficient to obtain the
explicit expressions for all helicity combinations of up to six gluons.

We will often use a more compact notation for the splitting
amplitude. We denote the splitting amplitude for $n$ collinear gluons,
of which $r$ have negative helicity, by:
\begin{eqnarray} \ssplit{1^+,\ldots , m_1^-, \ldots, m_2^-, \ldots,m_r^-,\ldots ,
  n^+}{\pm}=\Split_{\pm}(m_1,\ldots,m_r).
\end{eqnarray}

\subsubsection{$\Delta M = 0$}
\label{sec:4.1.1}

This is the simplest case which is read directly off the single MHV vertex.
The denominator of an $N$-point MHV amplitude is factorised as follows
(in the limit of collinear $p_1, \ldots, p_n$):
\bea
&&\spa{N,}.{1}\spa{1,}.{2}\ldots \spa{n,}.{n+1}\ldots \spa{N-1,}.{N} \nonumber \\
&&=
\Big(\sqrt{z_1 z_n} \prod_{l=1}^{n-1} \spa{l,}.{l+1}\Big) \times
\Big(\spa{N,}.{P} \spa{P,}.{n+1}\ldots \spa{N-1,}.{N} \Big)
\eea
where the first factor contributes to the splitting function, and the second one
is the denominator of the remaining hard MHV amplitude.
Hence, the splitting function is
\begin{eqnarray}
\ssplit{1^+ ,\ldots, n^+ }{+}= \frac{1}{\sqrt{z_1 z_n} \prod_{l=1}^{n-1} \spa{l,}.{l+1} } \ ,
\end{eqnarray}
and so by parity
\begin{eqnarray}
\ssplit{1^- ,\ldots , n^- }{-}= \frac{(-1)^{n-1}}{\sqrt{z_1 z_n} \prod_{l=1}^{n-1}
\spb{l,}.{l+1} } \ .
\end{eqnarray}
Similarly,
\begin{eqnarray}
\ssplit{1^+,\ldots,m_1^-, \ldots, n^+ }{-}= \frac{z_{m_1}^{2}}{\sqrt{z_1 z_n}
\prod_{l=1}^{n-1} \spa{l,}.{l+1} } \ ,
\end{eqnarray}
and
\begin{eqnarray}
\ssplit{1^-,\ldots,m_1^+, \ldots, n^- }{+}= \frac{(-1)^{n-1}z_{m_1}^{2}}{\sqrt{z_1 z_n}
\prod_{l=1}^{n-1} \spb{l,}.{l+1} } \ .
\end{eqnarray}

\subsubsection{$\Delta M = 1$}
\label{sec:4.1.2}

\begin{figure}[t!]
    \centering
    \psfrag{i+1+}{$~$}
    \psfrag{j+1+}{$~$}
    \psfrag{j+}{$j+$}
    \psfrag{i+}{{\small $i+$}}
    \psfrag{m1-}{$m_1^-$}
    \psfrag{m2-}{$m_2^-$}
    \begin{center}
        \includegraphics[width=5cm]{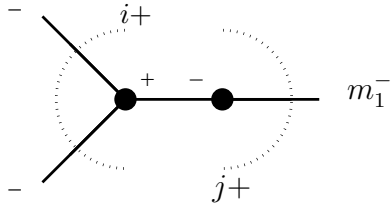}
    \end{center}
    \caption{MHV diagrams contributing to $\Split_+(m_1)$.  Negative helicity gluons are indicated
    by solid lines, while
    arbitrary numbers of positive helicity gluons emitted from each vertex are shown as dotted arcs.}
\label{fig:m1}
\end{figure}
This is the next-to-MHV (NMHV) case, and in the collinear limit we need to take
into account only a subset of MHV diagrams. In fact, there is only a single MHV
diagram (or more precisely a single class of MHV diagrams) which can contribute
to $\Split_+(m_1)$. It is shown in Fig.~\ref{fig:m1}.\footnote{MHV diagrams
where hard negative helicity gluons are emitted from more than one vertex do not
give rise to on-shell propagators and do not contribute
in the singular limit.} In the limit where gluons $1, \ldots, n$
become collinear.   The left vertex in Fig.~\ref{fig:m1} produces a `hard' MHV
amplitude while the right vertex generates the splitting function. We need to
sum over $i$ and $j$ in Fig.~\ref{fig:m1} in such a way that only diagrams with
a singular propagator are selected in the collinear limit. This puts a
constraint $j\le n$ where $n$ is the number of collinear gluons.  The resulting
splitting function reads,
\begin{align}
\label{mnpp}
\Split_+(m_1)
&= \frac{1}{\sqrt{z_1 z_n} \prod_{l=1}^{n-1} \spa{l,}.{l+1} }\bigg(
  \sum_{i=0}^{m_1-1} \sum_{j=m_1}^{n}
  \frac{\Del{i}.{j}.{m_1}^4}{D(i,j,q_{i+1,j})} \bigg)\ ,
\end{align}
where we define
\be
\label{D}
D(i,j,q)=\frac{q_{i+1,j}^2}{\spa{i,}.{i+1} \spa{j,}.{j+1}}
\Del{i}.{j}.{i}\Del{i}.{j}.{i+1}\Del{i}.{j}.{j}\Del{i}.{j}.{j+1}\ .
\ee

\begin{figure}[t!]
    \centering
    \psfrag{i+1+}{$~$}
    \psfrag{j+1+}{$~$}
    \psfrag{k+1+}{$~$}
    \psfrag{r+1+}{$~$}
    \psfrag{j+}{$j+$}
    \psfrag{i+}{$i+$}
    \psfrag{k+}{$k+$}
    \psfrag{r+}{$r+$}
    \psfrag{m1-}{$m_1^-$}
    \psfrag{m2-}{$m_2^-$}
    \begin{center}
        \includegraphics[width=12cm]{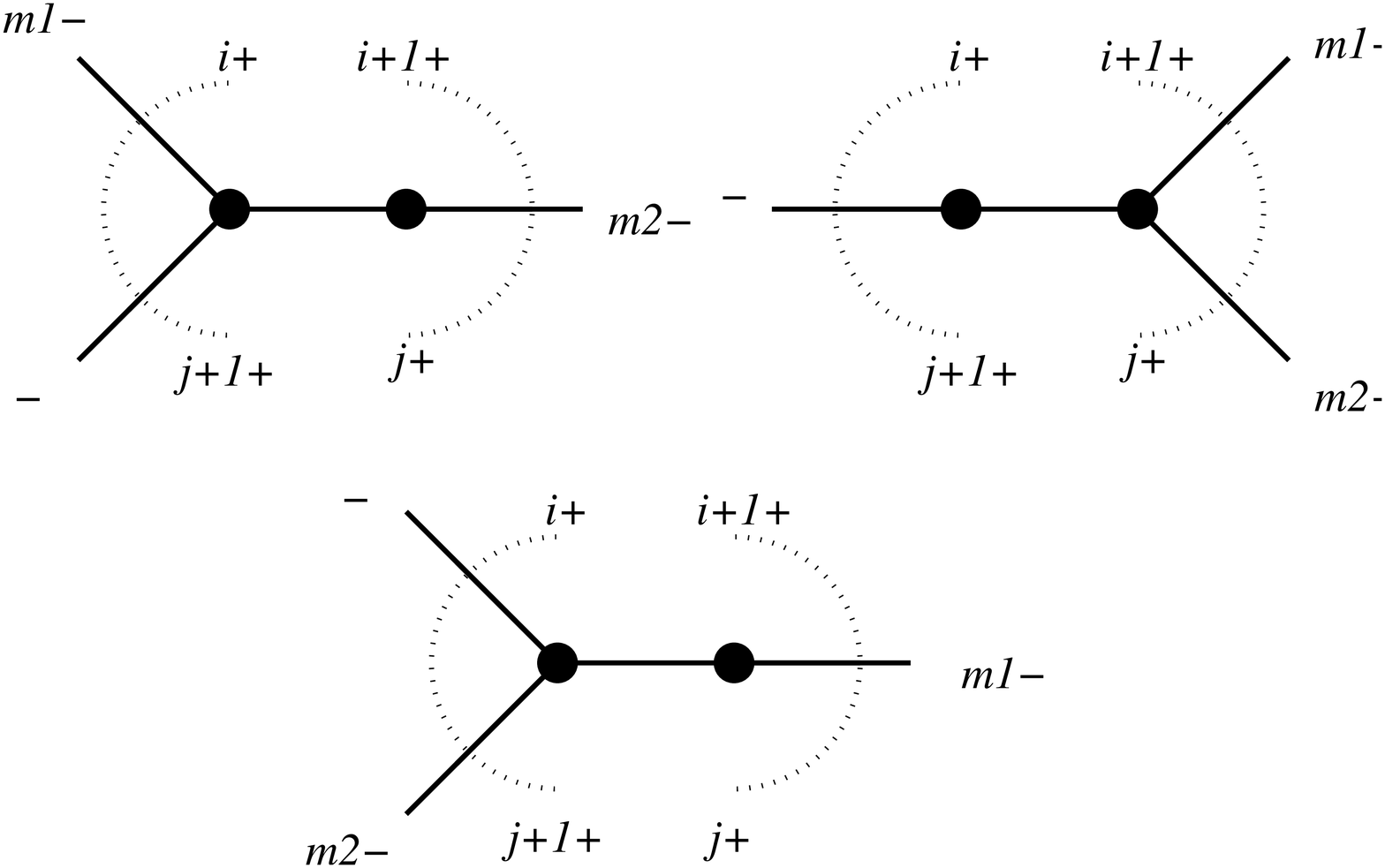}
    \end{center}
    \caption{MHV diagrams contributing to $\Split_-(m_1,m_2)$.}
\label{fig:m1m2}
\end{figure}

Similarly, there are three (classes of) MHV-diagrams contributing to $\Split_-(m_1,m_2)$.
They are shown in Fig.~\ref{fig:m1m2} and lead to a splitting function
which reads
\begin{align}
\label{mmnpm}
\Split_{-}(m_1,m_2)&=
 \frac{1}{\sqrt{z_1 z_n} \prod_{l=1}^{n-1} \spa{l,}.{l+1} }\Bigg(
  \sum_{i=0}^{m_1-1} \sum_{j=m_1}^{m_2-1} \frac{z_{m_2}^{2}\Del{i}.{j}.{m_1}^4}{D(i,j,q_{i+1,j})}
  \nonumber\\
  &+\sum_{i=m_1}^{m_2-1} \sum_{j=m_2}^{n}
  \frac{z_{m_1}^{2}\Del{i}.{j}.{m_2}^4}{D(i,j,q_{i+1,j})}\nonumber\\
  & +\sum_{i=0}^{m_1-1}
  \sum_{j=m_2}^{n} \frac{\spa{m_1}.{m_2}^4
   }{D(i,j,q_{i+1,j})} \Big( \sum_{{l=i+1}}^{{j}}z_l \Big)^4\Bigg)\ .
\end{align}

The remaining splitting amplitudes of the form
\begin{equation}
\ssplit{1^-, \ldots,
  m_1^+, \ldots, m_2^+, \ldots,m_r^+,\ldots, n^- }{\pm}
\end{equation}
are obtained by
parity transformation through the usual replacement $\spa{l,}.{k}
\leftrightarrow -\spb{l,}.{k}$.

\subsubsection{$\Delta M = 2$}
\label{sec:4.1.3}

\begin{figure}[t!]
    \centering
    \psfrag{i+1+}{$~$}
    \psfrag{j+1+}{$~$}
    \psfrag{k+1+}{$~$}
    \psfrag{r+1+}{$~$}
    \psfrag{j+}{$~$}
    \psfrag{i+}{$~$}
    \psfrag{k+}{$~$}
    \psfrag{r+}{$~$}
    \psfrag{m1-}{$m_1^-$}
    \psfrag{m2-}{$m_2^-$}
    \begin{center}
        \includegraphics[width=12cm]{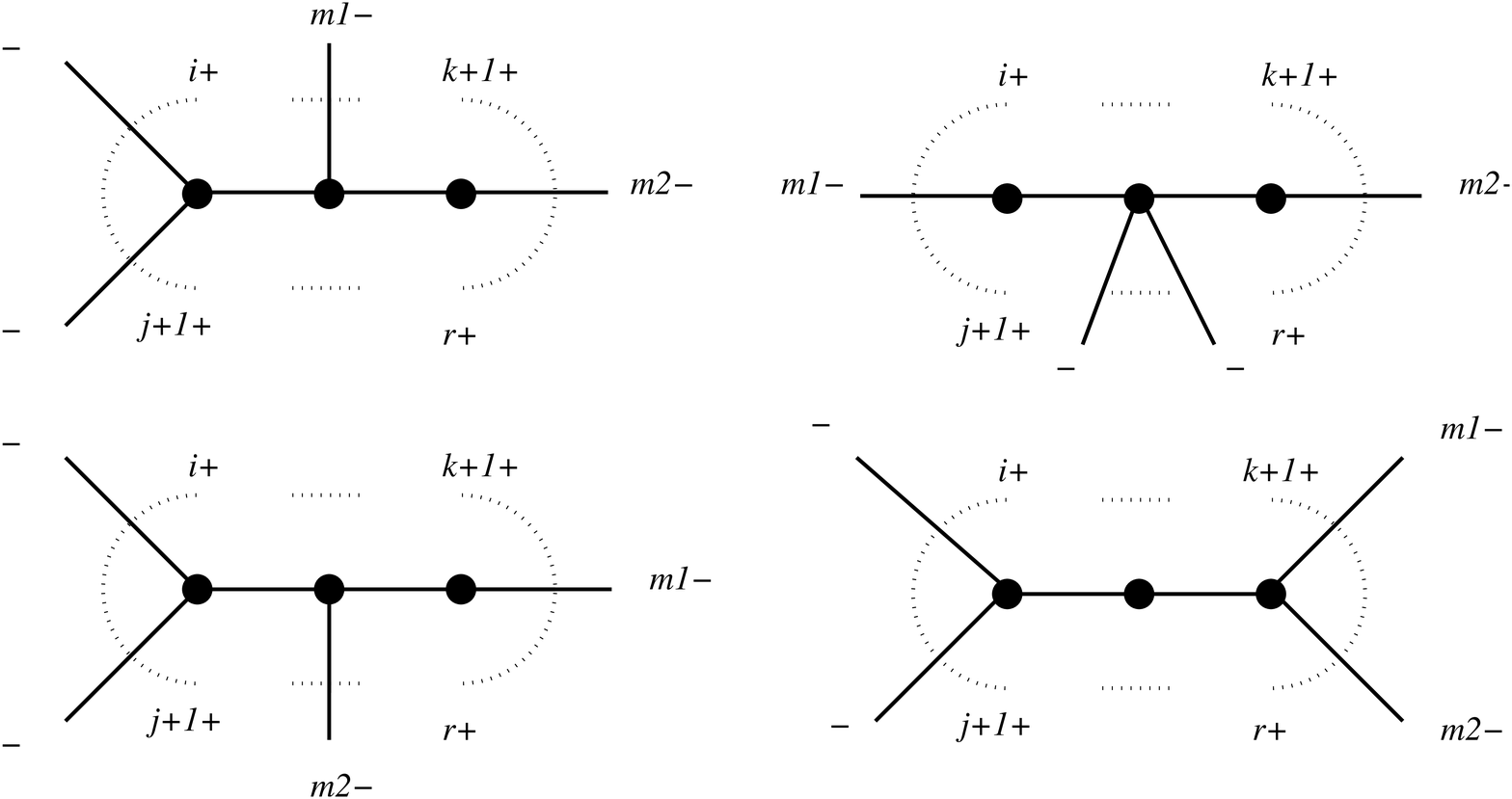}
    \end{center}
    \caption{MHV diagrams contributing to $\Split_+(m_1,m_2)$.}
\label{fig:m1m2plus}
\end{figure}

The collinear limits with $\Delta M = 2$ are derived from next-to-next-to-MHV
(NNMHV)
diagrams. There are four (classes of) MHV-diagrams contributing
to $\Split_{+}(m_1,m_2)$ which are shown in
Fig.~\ref{fig:m1m2plus}.  The corresponding splitting function is,
\begin{align}
\label{mmnpp}
\Split_{+}&(m_1,m_2)= \frac{1}{\sqrt{z_1 z_n} \prod_{l=1}^{n-1} \spa{l,}.{l+1}}\nonumber\\&
\Bigg( \sum_{i=0}^{m_1 -1}\sum_{j=m_2}^{n}\sum_{k=m_1}^{m_2-1}\sum_{r=m_2}^{j}
\frac{\Del{i}.{j}.{m_1}^4\Del{k}.{r}.{m_2}^4}{\DD(i,j,q_{i+1,j};k,r,q_{k+1,r})}\nonumber\\
&+ \sum_{i=0}^{m_1 -1}\sum_{j=m_1}^{k}\sum_{k=m_1}^{m_2-1}\sum_{r=m_2}^{n}
\frac{\Del{i}.{j}.{m_1}^4\Del{k}.{r}.{m_2}^4}{\DD(i,j,q_{i+1,j};k,r,q_{k+1,r})}\nonumber\\
& +\sum_{i=0}^{k}\sum_{j=m_2}^{n}\sum_{k=0}^{m_1-1}\sum_{r=m_1}^{m_2-1}
\frac{\Del{i}.{j}.{m_2}^4 \Del{k}.{r}.{m_1}^4}{\DD(i,j,q_{i+1,j};k,r,q_{k+1,r})}\nonumber\\
&+\sum_{i=0}^{k}\sum_{j=m_2}^{n}\sum_{k=0}^{m_1-1}\sum_{r=m_2}^{j}
\frac{\spa{m_1}.{m_2}^4 \Dell{i}.{j}.{k}.{r}^4}
{\DD(i,j,q_{i+1,j};k,r,q_{k+1,r})}
\Bigg)
\end{align}
where $\Del{i}.{j}.{k}$ is given in Eq. (\ref{eq:7}) and we introduce
\begin{equation}
  \label{eq:8}
\Dell{i}.{j}.{k}.{r} =
  \sum_{u=i+1}^{j}\sum_{v=k+1}^{r}\spa{u}.{v}\sqrt{z_u z_v} \ .
\end{equation}
The `effective propagator' $\DD$ is defined by
\begin{equation}
\DD(i,j,q_1;k,r,q_2)=\chi(i,k,q_1,q_2) \chi(r,j,q_2,q_1) \chi(j,k,q_1,q_2) 
D(i,j,q_1)D(k,r,q_2)
\end{equation}
in terms of $D$ defined previously in Eq.~\eqref{D}, and $\chi$ given by
\begin{equation}
 \label{chi}
\chi(i,k,q_1,q_2)= \left \{\begin{array} {ll} 1 & i \neq k \\
 \frac{\Dellq{1}.{2} \spa{i,}.{ i+1}}{  \Delq{ q_1}.{i+1}
 \Delq{q_2}.{i}  } & i = k
\end{array} \right.  .
\end{equation}

\begin{figure}[t!]
    \centering
    \psfrag{i+1+}{$(i+1)+$}
    \psfrag{j+1+}{$(j+1)+$}
    \psfrag{j+}{$j+$}
    \psfrag{i+}{$i+$}
    \psfrag{k+}{$k+$}
    \psfrag{r+}{$r+$}
    \psfrag{m1-}{$m_1^-$}
    \psfrag{m2-}{$m_2^-$}
    \begin{center}
        \includegraphics[width=12cm]{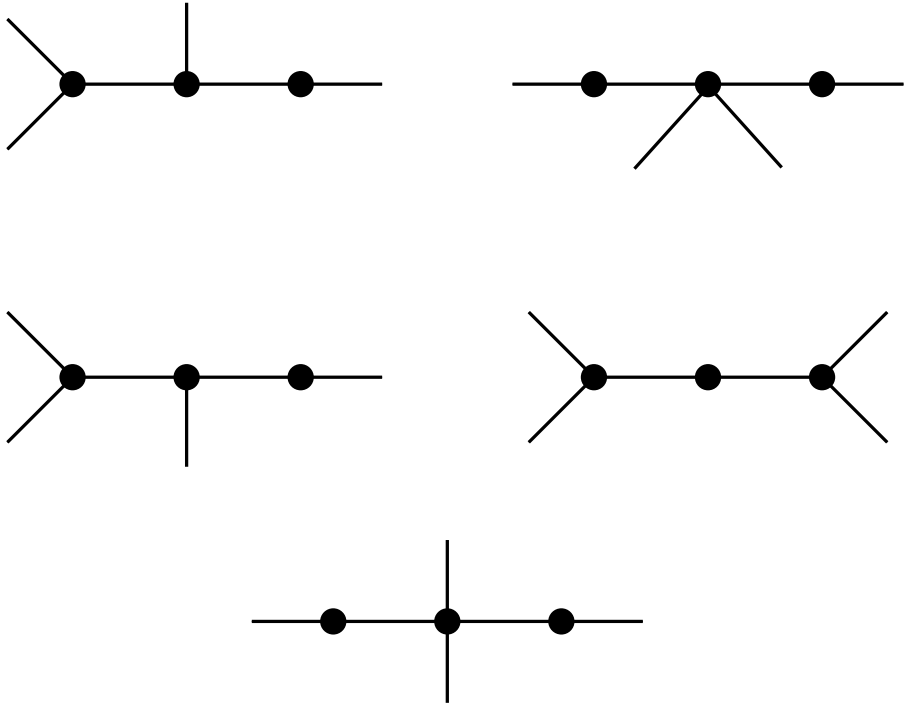}
    \end{center}
    \caption{MHV topologies contributing to $\Split_-(m_1,m_2,m_3)$.  
    The negative helicity gluons $m_1$, $m_2$ and $m_3$
    are distributed in a cyclic way around each diagram.  
    The remaining leg is the negative helicity gluon that
    remains after the collinear limit is taken.}
\label{fig:m1m2m3}
\end{figure}

Finally there are  16 classes of MHV-diagrams contributing to
$\Split_{-}(m_1,m_2,m_3)$,
coming from the 5 topologies shown in Fig.~\ref{fig:m1m2m3}
 and their cyclic permutations. The individual contributions are given by
\begin{eqnarray}
\label{mmmnpm}
\Split_{-}(m_1,m_2,m_3)=\frac{1}{\sqrt{z_1 z_n} \prod_{l=1}^{n-1}
  \spa{l,}.{l+1}}\sum_{i=1}^{16} A^{(i)}(m_1,m_2,m_3)
\end{eqnarray}
where
\begin{eqnarray*}
A^{(1)}(m_1,m_2,m_3)&=&\sum_{i=m_1}^{m_2-1}\sum_{j=m_3}^{n}\sum_{k=m_2}^{m_3-1}\sum_{r=m_3}^{j}
 z_{m_1}^2  \frac{\Del{i}.{j}.{m_2}^4 \Del{k}.{r}.{m_3}^4}{\DD(i,j,q_{i+1,j};k,r,q_{k+1,r})}
\\
A^{(2)}(m_1,m_2,m_3)&=&\sum_{i=m_1}^{k}\sum_{j=m_3}^{n}\sum_{k=m_1}^{m_2-1}\sum_{r=m_2}^{m_3-1}
 z_{m_1}^2  \frac{\Del{i}.{j}.{m_3}^4 \Del{k}.{r}.{m_2}^4}{\DD(i,j,q_{i+1,j};k,r,q_{k+1,r})}
\\
A^{(3)}(m_1,m_2,m_3)&=&\sum_{i=0}^{m_1-1}\sum_{j=m_2}^{m_3-1}\sum_{k=m_1}^{m_2-1}\sum_{r=m_2}^{j}
 z_{m_3}^2  \frac{\Del{i}.{j}.{m_1}^4 \Del{k}.{r}.{m_2}^4}{\DD(i,j,q_{i+1,j};k,r,q_{k+1,r})}
\\
A^{(4)}(m_1,m_2,m_3)&=&\sum_{i=0}^{k}\sum_{j=m_2}^{m_3-1}\sum_{k=0}^{m_1-1}\sum_{r=m_1}^{m_2-1}
 z_{m_3}^2  \frac{\Del{i}.{j}.{m_2}^4 \Del{k}.{r}.{m_1}^4}{\DD(i,j,q_{i+1,j};k,r,q_{k+1,r})}
\\
A^{(5)}(m_1,m_2,m_3)&=&\sum_{i=m_1}^{k}\sum_{j=m_3}^{n}\sum_{k=m_1}^{m_2-1}\sum_{r=m_3}^{j}
 z_{m_1}^2  \frac{\spa{m_2}.{m_3}^4 \Dell{i}.{j}.{k}.{r}^4}{\DD(i,j,q_{i+1,j};k,r,q_{k+1,r})}
\\
A^{(6)}(m_1,m_2,m_3)&=&\sum_{i=0}^{k}\sum_{j=m_2}^{m_3-1}\sum_{k=0}^{m_1-1}\sum_{r=m_2}^{j}
 z_{m_3}^2  \frac{\spa{m_1}.{m_2}^4 \Dell{i}.{j}.{k}.{r}^4}{\DD(i,j,q_{i+1,j};k,r,q_{k+1,r})}
\\
A^{(7)}(m_1,m_2,m_3)&=&\sum_{i=0}^{m_1-1}\sum_{j=m_3}^{n}\sum_{k=m_2}^{m_3-1}\sum_{r=m_3}^{j}
 \Big( \sum_{{l=i+1}}^{{j}}z_l \Big)^4   \frac{\spa{m_1}.{m_2}^4 \Del{k}.{r}.{m_3}^4}{\DD(i,j,q_{i+1,j};k,r,q_{k+1,r})}
\\
A^{(8)}(m_1,m_2,m_3)&=&\sum_{i=0}^{m_1-1}\sum_{j=m_3}^{n}\sum_{k=m_1}^{m_2-1}\sum_{r=m_3}^{j}
 \Big( \sum_{{l=i+1}}^{{j}}z_l \Big)^4  \frac{\spa{m_2}.{m_3}^4 \Del{k}.{r}.{m_1}^4}{\DD(i,j,q_{i+1,j};k,r,q_{k+1,r})}
\\
A^{(9)}(m_1,m_2,m_3)&=&\sum_{i=0}^{k}\sum_{j=m_3}^{n}\sum_{k=0}^{m_1-1}\sum_{r=m_2}^{m_3-1}
 \Big( \sum_{{l=i+1}}^{{j}}z_l \Big)^4  \frac{\spa{m_1}.{m_2}^4 \Del{k}.{r}.{m_3}^4}{\DD(i,j,q_{i+1,j};k,r,q_{k+1,r})}
\\
A^{(10)}(m_1,m_2,m_3)&=&\sum_{i=0}^{m_1-1}\sum_{j=m_3}^{n}\sum_{k=m_1}^{m_2-1}\sum_{r=m_2}^{m_3-1}
 \Big( \sum_{{l=i+1}}^{{j}}z_l \Big)^4  \frac{\spa{m_1}.{m_3}^4 \Del{k}.{r}.{m_2}^4}{\DD(i,j,q_{i+1,j};k,r,q_{k+1,r})}
\\
A^{(11)}(m_1,m_2,m_3)&=&\sum_{i=0}^{k}\sum_{j=m_3}^{n}\sum_{k=0}^{m_1-1}\sum_{r=m_1}^{m_2-1}
 \Big( \sum_{{l=i+1}}^{{j}}z_l \Big)^4  \frac{\spa{m_2}.{m_3}^4 \Del{k}.{r}.{m_1}^4}{\DD(i,j,q_{i+1,j};k,r,q_{k+1,r})}
\\
A^{(12)}(m_1,m_2,m_3)&=&\sum_{i=m_1}^{m_2-1}\sum_{j=m_3}^{n}\sum_{k=m_2}^{r}\sum_{r=m_2}^{m_3-1}
 z_{m_1}^2  \frac{\Del{i}.{k}.{m_2}^4 \Del{r}.{j}.{m_3}^4}{\DD(i,k,q_{i+1,k};r,j,q_{r+1,j})}
\\
A^{(13)}(m_1,m_2,m_3)&=&\sum_{i=0}^{m_1-1}\sum_{j=m_2}^{m_3-1}\sum_{k=m_1}^{r}\sum_{r=m_1}^{m_2-1}
 z_{m_3}^2  \frac{\Del{i}.{k}.{m_1}^4 \Del{r}.{j}.{m_2}^4}{\DD(i,k,q_{i+1,k};r,j,q_{r+1,j})}
\\
A^{(14)}(m_1,m_2,m_3)&=&\sum_{i=0}^{m_1-1}\sum_{j=m_3}^{n}\sum_{k=m_2}^{r}\sum_{r=m_2}^{m_3-1}
 \Big( \sum_{{l=i+1}}^{{k}}z_l \Big)^4  \frac{\spa{m_1}.{m_2}^4 \Del{r}.{j}.{m_3}^4}{\DD(i,k,q_{i+1,k};r,j,q_{r+1,j})}
\\
A^{(15)}(m_1,m_2,m_3)&=&\sum_{i=0}^{m_1-1}\sum_{j=m_3}^{n}\sum_{k=m_1}^{r}\sum_{r=m_1}^{m_2-1}
 \Big( \sum_{{l=r+1}}^{{j}}z_l \Big)^4  \frac{\spa{m_2}.{m_3}^4 \Del{i}.{k}.{m_1}^4}{\DD(i,k,q_{i+1,k};r,j,q_{r+1,j})}
\\
A^{(16)}(m_1,m_2,m_3)&=&\sum_{i=0}^{m_1-1}\sum_{j=m_3}^{n}\sum_{k=m_1}^{m_2-1}\sum_{r=m_2}^{m_3-1}
z_{m_2}^2  \frac{\Del{i}.{k}.{m_1}^4 \Del{r}.{j}.{m_3}^4}{\DD(i,k,q_{i+1,k};r,j,q_{r+1,j})}.
\end{eqnarray*}

\subsection{Specific results for $n<7$.}
\label{sec:4.2}

In this section we present compact expressions for splitting amplitudes
with up to six collinear gluons.
These results are obtained directly from
the general expressions given in Section \ref{sec:gen}.

First we note that
splitting amplitudes satisfy reflection symmetry,
\begin{align}
  \ssplit  {1^{\lambda_1},\ldots ,n^{\lambda_n}}{\pm} = (-1) ^ {n+1}
  \ssplit  {n^{\lambda_n},\ldots , 1^{\lambda_1}}{\pm}
\end{align}
and the dual Ward identity, see e.g. \cite{delduca},
\begin{align}
&\ssplit  {1^{\lambda_1},2^{\lambda_2},\ldots ,n^{\lambda_n}}{\pm}
+ \ssplit  {2^{\lambda_2},1^{\lambda_1},\ldots ,n^{\lambda_n}}{\pm}
+\cdots \nonumber \\
+\,\,&\ssplit  {2^{\lambda_2},\ldots ,1^{\lambda_1},n^{\lambda_n}}{\pm}
+ \ssplit  {2^{\lambda_2},\ldots ,n^{\lambda_n},1^{\lambda_1}}{\pm} = 0.
\end{align}
These relations reduce the number of independent splitting amplitudes significantly.

\subsubsection{$n=2$}
\label{sec:4.2.1}

For two collinear gluons there are two independent splitting
amplitudes with $\Delta M = 0$. All others can be obtained by parity and reflection.
Setting $z_1 = z$ and $z_2 = (1-z)$, we find
\begin{eqnarray}
\ssplit{1^+, 2^+}{+} &=& \, \frac{1}{\sqrt{z(1-z)}\,
\langle 1 2\rangle}\, ,\\
\ssplit{1^-, 2^+}{-} &=& \,\frac{z^2}{\sqrt{z(1-z)}\,
\langle 1 2\rangle}\, .
\end{eqnarray}
As expected, the splitting amplitudes have a single pole proportional to
$\spa{1}.{2}^{-1}$ .
Note that in the soft limit $z \to 0$, we see that helicity conservation
ensures that $\ssplit{1^-, 2^+}{-} \to 0$.

\subsubsection{$n=3$ result from MHV rules}
\label{sec:4.2.2}

For three collinear gluons there are three independent splitting
amplitudes with $\Delta M=0$. They all follow directly from a single MHV vertex
and are given by
\begin{eqnarray*}
  \label{eq:2}
 \ssplit {1^+,2^+,3^+}{+} &=&  \frac{1}{\sqrt{z_1z_3}\spa{1}.{2}\spa{2}.{3}}\ ,\\
  \ssplit{1^-,2^+,3^+}{-} &=& \frac{z_1^2}{\sqrt{z_1z_3}\spa{1}.{2}\spa{2}.{3}}\ ,\\
  \ssplit{1^+,2^-,3^+}{-} &=& \frac{z_2^2}{\sqrt{z_1z_3}\spa{1}.{2}\spa{2}.{3}}\ .
\end{eqnarray*}
Parity and the reflection symmetry,
$\ssplit{1^+,2^+,3^-}{-} = \ssplit{3^-,2^+,1^+}{-},$
give the rest.

When $\Delta M = 1$, there are three amplitudes,
\bea
\label{eq:triple1}
\ssplit{1^-, 2^+, 3^+}{+} &=& {\frac {\spa{1}.{2}{z_{{2}}}^{2}}
{\sqrt {z_{{1}}z_{{2}}z_{{3}}}s_{{1,2}}
 \left( z_{{1}}+z_{{2}} \right)  \left( \spa{1}.{3}\sqrt {z_{{1}}}+\spa{2}.{3}\sqrt {z_{{2}}}
 \right) }}
 \nonumber\\
&+&
{\frac { \left( \spa{1}.{2}\sqrt {z_{{2}}}+\spa{1}.{3}\sqrt {z_{{3}}} \right) ^{3}}
 {s_{{1,3}} \spa{1}.{2}\spa{2}.{3} \left( \spa{1}.{3}\sqrt {z_{{1}}}+\spa{2}.{3}\sqrt {z_{{2}}}
  \right) }} \ , \\
  \nonumber\\
\label{eq:triple3}
\ssplit{1^+, 2^-, 3^+}{+} &=& -\ssplit{2^-, 1^+, 3^+}{+}-\ssplit{1^+, 3^+, 2^-}{+}\nonumber \\
&=&
{\frac {\spa{1}.{2}{z_{{1}}}^{2}}{\sqrt {z_{{1}}z_{{2}}z_{{3}}}s_{{1,2}}
 \left( z_{{1}}+z_{{2}} \right)  \left( \spa{1}.{3}
\sqrt {z_{{1}}}+\spa{2}.{3}\sqrt {z_{{2}}} \right) }}\nonumber\\
&+& {\frac { \left(
\spa{2}.{1}\sqrt {z_{{1}}}+\spa{2}.{3}\sqrt {z_{{3}}} \right) ^{4}}{
s_{{1,3}}\spa{1}.{2} \spa{2}.{3}
\left( \spa{1}.{3}\sqrt {z_{{1}}}+\spa{2}.{3}\sqrt {z_{{2}}} \right)
\left( \spa{1}.{2}\sqrt {z_{{2}}}+\spa{1}.{3}\sqrt {z_{{3}}} \right)
}}\nonumber\\
&+& {\frac {\spa{2}.{3}{z_{{3}}}^{2}
}{\sqrt {z_{{1}}z_{{2}}z_{{3}}}s_{{2,3}} \left( z_{{2}}+z_{{3}} \right) \left( \spa{1}.{2}
\sqrt {z_{{2}}}+
\spa{1}.{3}\sqrt {z_{{3}}} \right)  }} \ , \\
\nonumber\\
\ssplit{1^+, 2^+, 3^-}{+} &=& \ssplit{3^- ,2^+, 1^+}{+} \ .
\\ \nonumber
\eea
In addition to singular terms like $\spa{1}.{2}$, we see that the splitting functions
contain mixed terms like $s_{1,3}$.  The net singularity is schematically of the form
$[~]\langle~\rangle$.

Note that $\ssplit{1^-, 2^+, 3^+}{+}$ contains poles in $s_{1,2}$ and the triple invariant
$s_{1,3} = s_{123}$ but not in $s_{2,3}$.  This
is because there is no MHV rule graph with a three-point vertex involving two positive helicity
gluons.

Expressions for these splitting functions are given in Eq.~(5.52) of
Ref.~\cite{delduca}. The results given here are more compact and have a
rather different analytic form. After adjusting the
normalisation of the colour matrices, the splitting functions of
Eqs.~\eqref{eq:triple1}--\eqref{eq:triple3} numerically agree with those of
Ref.~\cite{delduca}.  

\subsubsection{$n=3$ result from the BCF recursion relation}
\label{sec:4.2.3}

We now want to rederive the above results using the BCF recursion relation of
\cite{BCF4}. In doing this we will (a) draw some useful comparisons
between the `BCF recursion' and the `MHV rules' formalisms
from the perspective of collinear amplitudes; and (b) test our
expressions, such as Eq.~\eqref{eq:triple1} for $\ssplit{1^-, 2^+, 3^+}{+}$.

\begin{figure}[t!]
    \centering
    \psfrag{1-}{$\widehat 1^-$}
    \psfrag{2+}{$\widehat 2^+$}
    \psfrag{3+}{$3^+$}
    \psfrag{4+}{$4^+$}
    \psfrag{5+}{$5^+$}
    \psfrag{6-}{$6^-$}
    \psfrag{+}{$+$}
    \psfrag{-}{$-$}
    \begin{center}
        \includegraphics[height=4cm]{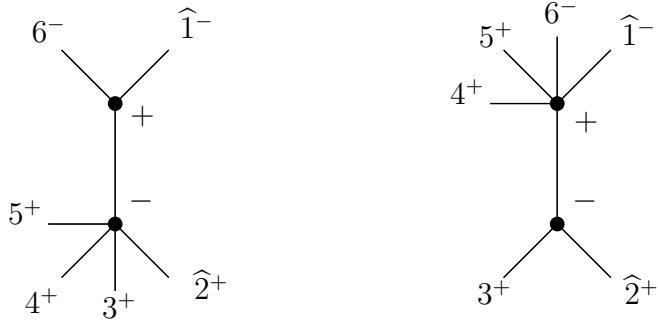}
    \end{center}
    \caption{BCF diagrams contributing to $A(1^-,2^+,3^+,4^+,5^-,6^-)$.}
\label{fig:bcf}
\end{figure}

We start with the six-point amplitude $A(1^-,2^+,3^+,4^+,5^-,6^-)$, and
calculate it via the BCF recursive approach. We ultimately want to take
the collinear limit $1\,||\,2\,||\,3 \rightarrow P^{+}$, so it will be
convenient to choose the `marked' gluons (required for the BCF recursive set-up)
to be from this collinear set. Hence, we will mark the
$\hat{1}^-$ and $\hat{2}^+$ gluons. There are only two BCF diagrams
which contribute to the full amplitude, and they are shown in Fig.~\ref{fig:bcf}.
We now note that in this particular collinear limit, only the second of these diagrams
contains an on-shell propagator, $1/s_{23}$. Nevertheless, in distinction
with the MHV rules approach which we have adopted previously,
both BCF diagrams need to be taken into account in the collinear limit.

The full amplitude reads
\begin{equation}
  A(1^-,2^+,3^+,4^+,5^-,6^-) = \frac{1}{\asb{3}{1+2}{6}}\left(
    \frac{\asb{5}{6+1}{2}^3}{\ssb{6}{1}\ssb{1}{2}\aab{3}{4}\aab{4}{5}s_{3,5}}
    + \frac{\asb{1}{2+3}{4}^3}{\ssb{4}{5}\ssb{5}{6}\aab{1}{2}\aab{2}{3}s_{1,3}}
\right) \ ,
\label{eq:1check}
\end{equation}
where the two terms on the right hand side correspond to the two BCF diagrams above
(cf. Eq.~(2.9) of Ref.~\cite{BCF4}).

In the $1\,||\,2\,||\,3 \rightarrow P^{+} $ collinear limit, the first term becomes
\bea
 \frac{\aab{1}{2}z_2^2 }{\sqrt{z_1 z_2 z_3}s_{1,2} (z_1+z_2) (\aab{1}{3}\sqrt{z_1}+\aab{2}{3}
 \sqrt{z_2})}
 \times
 \frac{\aab{5}{6}^4}{\aab{5}{6}\aab{6}{P}\aab{P}{4}\aab{4}{5}}.
\eea
This term factors into a contribution to the splitting amplitudes multiplied by
a four-point MHV vertex.
In contrast, in the collinear limit the second term factors onto the $\overline{{\rm MHV}}$
type diagram, written in terms of the
anti-holomorphic spinor products,
\bea
 \frac{ (\aab{1}{2}\sqrt{z_2}+\aab{1}{3}\sqrt{z_3})^3}{ s_{1,3} \aab{1}{2} \aab{2}{3}
 (\aab{1}{3}\sqrt{z_1}+\aab{2}{3}\sqrt{z_2})}
\times
 \frac{\ssb{P}{4}^4}{\ssb{P}{4}\ssb{4}{5}\ssb{5}{6}\ssb{6}{P}}.
\eea
For the special case of four-point amplitudes, the $\overline{{\rm MHV}}$
and MHV amplitudes coincide
and we find an identical result to Eq.~\eqref{eq:triple1}.

Likewise, to test our expression for $\ssplit{1^+, 2^-, 3^+}{+}$
we start from Eq.~(3.4) in \cite{BCF4};
\begin{eqnarray}
  A(1^+,2^-,3^+,4^-,5^+,6^-)&=& \frac{\ssb{1}{3}^4 \aab{4}{6}^4
  }{\ssb{1}{2}\ssb{2}{3}\aab{4}{5}\aab{5}{6}
  s_{1,3}\asb{6}{1+2}{3}\asb{4}{2+3}{1}}\nonumber\\
&+&\frac{\aab{2}{6}^4 \ssb{3}{5}^4 }{\aab{6}{1}\aab{1}{2}\ssb{3}{4}\ssb{4}{5}
  s_{3,5}\asb{6}{4+5}{3}\asb{2}{3+4}{5}}\nonumber\\
&+&\frac{\ssb{1}{5}^4 \aab{2}{4}^4 }{\aab{2}{3}\aab{3}{4}\ssb{5}{6}\ssb{6}{1}
  s_{2,4}\asb{4}{2+3}{1}\asb{2}{3+4}{5}}.
\label{eq:3check}
\end{eqnarray}
Taking the collinear limit $1\,||\,2\,||\,3 \rightarrow P^{+} $, we find that
\begin{eqnarray}
  \label{eq:4check}
  \mbox{Split}(1^+,2^-,3^+\to P^+)
&=&\frac{z_2^2 z_3^2  \ssb{1}{2}}{\sqrt{z_1 z_2 z_3} s_{1,2}
  (z_1+z_2)\left(\ssb{1}{3}\sqrt{z_1}+\ssb{2}{3}\sqrt{z_2}\right)}\nonumber\\
&+&
  \frac{\ssb{1}{3}^4}{s_{1,3}\ssb{1}{2}\ssb{2}{3}\left(\ssb{1}{3}\sqrt{z_1}
  +\ssb{2}{3}\sqrt{z_2}\right)\left(\ssb{1}{2}\sqrt{z_2}+\ssb{1}{3}\sqrt{z_3}\right)}\nonumber\\
&+&\frac{z_1^2 z_2^2 \ssb{2}{3}}{\sqrt{z_1 z_2 z_3}s_{2,3}(z_2+z_3)
\left(\ssb{1}{2}\sqrt{z_2}+\ssb{1}{3}\sqrt{z_3}\right)}.
\end{eqnarray}
This result has the same kinematic-invariant
pole structure as Eq.~\eqref{eq:triple3}, but otherwise is not obviously equivalent to
Eq.~\eqref{eq:triple3}.
Note that
Eq.~\eqref{eq:4check} contains terms like
$\left(\ssb{1}{2}\sqrt{z_2}+\ssb{1}{3}\sqrt{z_3}\right)$ (rather than
$\left(\aab{1}{2}\sqrt{z_2}+\aab{1}{3}\sqrt{z_3}\right)$).
Despite appearances, a more careful (e.g. numerical)
comparison shows that these two results, Eqs.~\eqref{eq:triple3}
and \eqref{eq:4check}, are in fact the same.

\subsubsection{$n=4$}
\label{sec:4.2.4}

For $n=4$, there are five collinear limits coming directly
from MHV amplitudes
where the number of gluons with negative helicity doesn't change, $\Delta M=0$,
\begin{eqnarray}
\ssplit{1^+ ,2^+, 3^+, 4^+}{+}&=& \frac{1}{\sqrt{z_1z_4}\spa{1}.{2}
  \spa{2}.{3} \spa{3}.{4}},\\
\ssplit{1^- ,2^+, 3^+, 4^+}{-}&=& \frac{z_1^2}{\sqrt{z_1z_4}\spa{1}.{2}
  \spa{2}.{3} \spa{3}.{4}},\\
\ssplit{1^+ ,2^-, 3^+, 4^+}{-}&=& \frac{z_2^2}{\sqrt{z_1z_4}\spa{1}.{2}
  \spa{2}.{3} \spa{3}.{4}}.
\end{eqnarray}
The remaining two are obtained by reflection symmetry,
\begin{eqnarray}
\ssplit{1^+, 2^+, 3^-, 4^+}{-}&=&-\ssplit{4^+ ,3^-, 2^+ ,1^+}{-}, \\
\ssplit{1^+ ,2^+ ,3^+ ,4^-}{-}&=&-\ssplit{4^- ,3^+ ,2^+, 1^+}{-}.
\end{eqnarray}

When $\Delta M = 1$,  there are ten splitting amplitudes however only three
are imdependent, 
\begin{eqnarray}
\lefteqn{\ssplit{1^- ,2^+, 3^+, 4^+}{+}=\mathcal{B}_1(1,2,3,4)} \nonumber \\
&=&{
-\frac { z_2^{3/2}\spa{1}.{2}}
{\sqrt {z_1z_{{4}}}\spa{3}.{4}s_{{1,2}} \left( z_{{1}}+z_{{2}} \right) 
\Del{0}.{2}.{3}}}\nonumber\\&&
+{\frac {  \Del{0}.{3}.{1}  ^{3}}
{\sqrt {z_{{4}}}\spa{1}.{2}\spa{2}.{3}s_{{1,3}} 
\left( z_{{1}}+z_{{2}}+z_{{3}} \right) \Del{0}.{3}.{3}\Del{0}.{3}.{4}}}\nonumber\\&&
-{\frac { \Del{0}.{4}.{1} ^{3}}{\spa{
1}.{2}\spa{2}.{3}\spa{3}.{4}s_{{1,4}}\Del{0}.{4}.{4}}},
\label{eq:spb1}
\end{eqnarray}
\begin{eqnarray}
 \lefteqn{ \ssplit{1^-,2^-,3^+,4^+}{-}=\mathcal{B}_2(1,2,3,4) }\nonumber \\
&=&
-{\frac {z_1^{3/2} z_3^{3/2} \spa{2}.{3}}{
	\sqrt {z_2z_{{4}}}s_{{2,3}}\Del{1}.{3}.{1}\Del{1}.{
        3}.{4}}}\nonumber\\&&-{\frac {{z_{{1}}}^{3/2}  
        \Del{1}.{4}.{2}  ^{3} }{\spa{2}.{3}\spa{3}.{4}s_{{2,4}}
      \Del{1}.{4}.{1}\Del{1}.{4}.{4} \left( 1-z_{{1}}
      \right) }}\nonumber\\&&-{\frac {  \spa{1}.{2}
       \left( z_{{1}}+z_{{2}} \right) ^{3}}
       {\sqrt {z_1z_2z_{{4}}}\spa{3}.{4}s_{{1,2}}\Del{0}.{2}.{3}}}\nonumber\\
       &&+{\frac
    {  \spa {1}.{2} ^{3} \left( 1-z_{{4}}
      \right) ^{3}}{\sqrt
      {z_{{4}}}\spa{2}.{3}s_{{1,3}}\Del{0}.{3}.{1}
      \Del{0}.{3}.{3}\Del{0}.{3}.{4}}}\nonumber\\&&-{\frac
    {  \spa{1}.{2}  ^
      {3}}{s_{{1,4}}\Del{0}.{4}.{1}\Del{0}.{4}.{4}\spa{2}.{3}\spa{3}.{4}}},
\label{eq:spb2}
\end{eqnarray}
and,
\begin{eqnarray}
  \lefteqn{\ssplit{1^-,2^+,3^-,4^+}{-}=\mathcal{B}_3(1,2,3,4)} \nonumber \\
&=&-{\frac {z_2^{3/2}{z_{{3}}}^{2}  
        \spa{1}.{2}}{\sqrt{z_{{1}}z_{{4}}}\spa{3}.{4}s_{{1,2}} \left( z_{{1
          }}+z_{{2}} \right)
      \Del{0}.{2}.{3}}}\nonumber\\&&-{\frac
    {z_1^{3/2}z_2^{3/2} \spa{2}.{3}}{\sqrt
      {z_{{3}}z_{{4}}}s_{{2,3}}\Del{1}.{3}.{1}\Del{1}.{3}.{4}}}\nonumber\\&&-{\frac
    {{ z_{{1}}}^{3/2}  \Del{1}.{4}.{3}   ^{4}}{\spa{2}.{3}\spa{3}.{4}s_{{2,4}}\Del{1}.{4}.{1}
      \Del{1}.{4}.{2}\Del{1}.{4}.{4} \left( 1-z_1
      \right) }}\nonumber\\&&-{\frac {{z_{{1}}}^{3/2}  
        \Del{2}.{4}.{3}  ^{3}}{\spa{1}.{2}\spa{3}.{4}s_{{3,4}}\Del{2}.{4
      }.{2}\Del{2}.{4}.{4} \left( z_{{3}}+z_{{4}} \right)
    }}\nonumber\\&&+{\frac { \spa{1}.{3} 
      ^{4} \left( z_{{1}}+z_{{2}}+z_{{3}} \right) ^{3}}{\sqrt {z_{{4}}}\spa{1}.{2}\spa{2}
      .{3}s_{{1,3}}\Del{0}.{3}.{1}\Del{0}.{3}.{3}\Del{0}.{3}.{4}}}\nonumber\\&&-{\frac
    { \spa{1}.{3} 
      ^{4}}{\spa{1}.{2}\spa{2}.{3}\spa{3}.{4}s_{{1,4}}
      \Del{0}.{4}.{4}\Del{0}.{4}.{1}}},
\label{eq:spb3}
\end{eqnarray}
where $\Del{i}.{j}.{k}$ is given in Eq.~\eqref{eq:7}.
The seven remaining $\Delta M=1$ splitting functions can be obtained by 
using the dual ward identity,
\begin{eqnarray}
\ssplit{1^+ ,2^- , 3^+ ,4^+}{+}&=&
-\mathcal{B}_1(2,1,3,4)-\mathcal{B}_1(2,3,1,4)-\mathcal{B}_1(2,3,4,1),\nonumber \\
\ssplit{1^+, 2^+, 3^-, 4^+}{+}&=& \phantom{-}
\mathcal{B}_1(3,4,2,1)+\mathcal{B}_1(3,2,4,1)+\mathcal{B}_1(3,2,1,4),\nonumber \\
\ssplit{1^-, 2^+, 3^+,4^-}{-}&=& \phantom{-}\mathcal{B}_3(4,3,1,2) +
\mathcal{B}_2(4,1,3,2) + \mathcal{B}_2(1,4,3,2) ,\nonumber \\
\ssplit{1^+, 2^-,  3^-, 4^+}{-}&=& - \mathcal{B}_3(2,1,3,4) -
\mathcal{B}_2(2,3,1,4) - \mathcal{B}_2(2,3,4,1),\nonumber \\
\end{eqnarray}
or
reflection symmetry, 
\begin{eqnarray}
\ssplit{1^+ ,2^+ ,3^+ ,4^-}{+}&=& - \ssplit{4^- ,3^+, 2^+, 1^+}{-},\nonumber \\
 \ssplit{1^+,2^-, 3^+,4^-}{-}&=&- \ssplit{4^- ,3^+, 2^-, 1^+}{-},\nonumber \\
 \ssplit{1^+,2^+ ,3^-,4^-}{-}&=&- \ssplit{4^-,3^- ,2^+,1^+}{-}.
\end{eqnarray}

Finally splitting functions with $\Delta M = 2,~3$ are related to those given
above by the parity transformation.

Inspection of Eqs.~\eqref{eq:spb1}, \eqref{eq:spb2} and \eqref{eq:spb3} reveals
that each term is inversely proportional to a single invariant, in keeping with
its MHV rules origins.   For this type of collinear limit, there are potentially
six invariants, the double invariants $s_{1,2}, s_{2,3}, s_{3,4}$, 
the triple invariants $s_{1,3}, s_{2,4}$ and $s_{1,4}$.  Some poles are absent
because the MHV rules forbid that type of contribution.  For example, in 
$\ssplit{1^- ,2^+, 3^+, 4^+}{+}$, there are no contributions with poles in
$s_{2,3}$, $s_{3,4}$ or $s_{2,4}$ precisely because these poles correspond to
forbidden MHV diagrams.

Expressions for the four gluon splitting functions are given in 
Ref.~\cite{delduca}. The results given here are more compact and have a
rather different analytic form. After adjusting the
normalisation of the colour matrices, the splitting functions of
Eqs.~\eqref{eq:spb1}--\eqref{eq:spb3} numerically agree with those of
Ref.~\cite{delduca}.  

\subsubsection{$n=5$}
\label{sec:4.2.5}

In total there are 64 different splitting amplitudes, but only eleven are
independent. The rest can be obtained with the help of parity,
reflection and dual ward identities.
The three
simplest independent collinear limits can be obtained using only MHV rules,
\begin{eqnarray*}
\ssplit{1^+ ,2^+, 3^+, 4^+, 5^+}{+}&=& \frac{1}{\sqrt{z_1z_5}\spa{1}.{2}
  \spa{2}.{3}  \spa{3}.{4} \spa{4}.{5}},\\
\ssplit{1^-, 2^+,  3^+, 4^+, 5^+}{-} &=& \frac{z_1^2}{\sqrt{z_1z_5}\spa{1}.{2}
  \spa{2}.{3}  \spa{3}.{4} \spa{4}.{5}}, \\
\ssplit{1^+, 2^-,  3^+, 4^+, 5^+}{-}&=& \frac{z_2^2}{\sqrt{z_1z_5}\spa{1}.{2}
  \spa{2}.{3}  \spa{3}.{4} \spa{4}.{5}}.
\end{eqnarray*}
The amplitudes with $\Delta M = 1$ require the application of Eqs.~(\ref{mnpp}) and
(\ref{mmnpm}). There are 5 independent amplitudes in this class of splitting
amplitudes, but we give here only two examples, one for each of the
cases $-\to +$ and $-- \to -$,
\begin{align}
&\ssplit{1^-, 2^+,  3^+, 4^+, 5^+}{+}=
\nonumber\\
&= {\frac { \left( \Del{0}.{2}.{1} \right) ^{3}\sqrt {z_{{1}}}}{\sqrt {z_
{{1}}z_{{5}}}\spa{1}.{2}\spa{3}.{4}\spa{4}.{5}s_{{1,2}} \left( z_{{1}}
+z_{{2}} \right) \Del{0}.{2}.{2}\Del{0}.{2}.{3}}}\nonumber\\
&+{\frac { \left( \Del
{0}.{3}.{1} \right) ^{3}\sqrt {z_{{1}}}}{\sqrt {z_{{1}}z_{{5}}}\spa{1}
.{2}\spa{2}.{3}\spa{4}.{5}s_{{1,3}} \left( z_{{1}}+z_{{2}}+z_{{3}}
 \right) \Del{0}.{3}.{3}\Del{0}.{3}.{4}}}\nonumber\\
&+{\frac { \left( \Del{0}.{4}.
{1} \right) ^{3}\sqrt {z_{{1}}}}{\sqrt {z_{{1}}z_{{5}}}\spa{1}.{2}\spa
{2}.{3}\spa{3}.{4}s_{{1,4}} \left( z_{{1}}+z_{{2}}+z_{{3}}+z_{{4}}
 \right) \Del{0}.{4}.{4}\Del{0}.{4}.{5}}}\nonumber\\
&-{\frac { \left( \Del{0}.{5}.
{1} \right) ^{3}}{\spa{1}.{2}\spa{2}.{3}\spa{3}.{4}\spa{4}.{5}s_{{1,5}
}\Del{0}.{5}.{5}}},
\end{align}
and,
\begin{align}
&\ssplit{1^-, 2^-,  3^+, 4^+, 5^+}{-}= \nonumber\\
& {\frac {{z_{{1}}}^{2} \left( \Del{1}.{3}.{2} \right) ^{3}}{\sqrt {z_{{
1}}z_{{5}}}\spa{2}.{3}\spa{4}.{5}s_{{2,3}}\Del{1}.{3}.{1}\Del{1}.{3}.{
3}\Del{1}.{3}.{4}}}\nonumber\\&+{\frac {{z_{{1}}}^{2} \left( \Del{1}.{4}.{2}
 \right) ^{3}}{\sqrt {z_{{1}}z_{{5}}}\spa{2}.{3}\spa{3}.{4}s_{{2,4}}
\Del{1}.{4}.{1}\Del{1}.{4}.{4}\Del{1}.{4}.{5}}}\nonumber\\&-{\frac {{z_{{1}}}^{2}
 \left( \Del{1}.{5}.{2} \right) ^{3}\sqrt {z_{{5}}}}{\sqrt {z_{{1}}z_{
{5}}}\spa{2}.{3}\spa{3}.{4}\spa{4}.{5}s_{{2,5}}\Del{1}.{5}.{1}\Del{1}.
{5}.{5} \left( z_{{2}}+z_{{3}}+z_{{4}}+z_{{5}} \right) }}\nonumber\\&+{\frac {
 \left( \spa{1}.{2} \right) ^{3} \left( z_{{1}}+z_{{2}} \right) ^{3}
\sqrt {z_{{1}}}}{\sqrt {z_{{1}}z_{{5}}}\spa{3}.{4}\spa{4}.{5}s_{{1,2}}
\Del{0}.{2}.{1}\Del{0}.{2}.{2}\Del{0}.{2}.{3}}}\nonumber\\&+{\frac { \left( \spa{1
}.{2} \right) ^{3} \left( z_{{1}}+z_{{2}}+z_{{3}} \right) ^{3}\sqrt {z
_{{1}}}}{\sqrt {z_{{1}}z_{{5}}}\spa{2}.{3}\spa{4}.{5}s_{{1,3}}\Del{0}.
{3}.{1}\Del{0}.{3}.{3}\Del{0}.{3}.{4}}}\nonumber\\&+{\frac { \left( \spa{1}.{2}
 \right) ^{3} \left( z_{{1}}+z_{{2}}+z_{{3}}+z_{{4}} \right) ^{3}
\sqrt {z_{{1}}}}{\sqrt {z_{{1}}z_{{5}}}\spa{2}.{3}\spa{3}.{4}s_{{1,4}}
\Del{0}.{4}.{1}\Del{0}.{4}.{4}\Del{0}.{4}.{5}}}\nonumber\\&-{\frac { \left( \spa{1
}.{2} \right) ^{3}}{\spa{2}.{3}\spa{3}.{4}\spa{4}.{5}s_{{1,5}}\Del{0}.
{5}.{1}\Del{0}.{5}.{5}}}.
\end{align}

The most complicated amplitudes are those with $\Delta M = 2 $ and
require the use of Eq.~(\ref{mmnpp}). There are three  independent
splitting functions, but here we only give one example.
\begin{align}
&\ssplit{1^-, 2^-,  3^+, 4^+, 5^+}{+}=
\nonumber\\&
-{\frac { \left( \Del{0}.{3}.{1} \right) ^{3} \left( \Del{1}.{3}.{2}
 \right) ^{3}}{\sqrt {z_{{5}}}\spa{2}.{3}\spa{4}.{5}\Dell{0}.{3}.{1}.{
3}s_{{1,3}} \left( z_{{1}}+z_{{2}}+z_{{3}} \right) \Del{0}.{3}.{4}s_{{
2,3}}\Del{1}.{3}.{1}\Del{1}.{3}.{3}}}\nonumber\\&+{\frac { \left( \Del{0}.{4}.{1}
 \right) ^{3} \left( \Del{1}.{3}.{2} \right) ^{3}}{\sqrt {z_{{5}}}\spa
{2}.{3}s_{{1,4}} \left( 1-z_5 \right) \Del{0
}.{4}.{4}\Del{0}.{4}.{5}s_{{2,3}}\Del{1}.{3}.{1}\Del{1}.{3}.{3}\Del{1}
.{3}.{4}}}\nonumber\\&-{\frac { \left( \Del{0}.{4}.{1} \right) ^{3} \left( \Del{1}
.{4}.{2} \right) ^{3}}{\sqrt {z_{{5}}}\spa{2}.{3}\spa{3}.{4}\Dell{0}.{
4}.{1}.{4}s_{{1,4}} \left( 1 -z_5 \right)
\Del{0}.{4}.{5}s_{{2,4}}\Del{1}.{4}.{1}\Del{1}.{4}.{4}}}\nonumber\\&-{\frac {
 \left( \Del{0}.{5}.{1} \right) ^{3} \left( \Del{1}.{3}.{2} \right) ^{
3}}{\Del{1}.{3}.{4}\spa{2}.{3}\spa{4}.{5}s_{{1,5}}\Del{0}.{5}.{5}\Del{
1}.{3}.{1}\Del{1}.{3}.{3}s_{{2,3}}}}\nonumber\\&-{\frac { \left( \Del{1}.{4}.{2}
 \right) ^{3} \left( \Del{0}.{5}.{1} \right) ^{3}}{s_{{2,4}}\Del{1}.{4
}.{1}\Del{1}.{4}.{4}\Del{1}.{4}.{5}\spa{2}.{3}\spa{3}.{4}s_{{1,5}}\Del
{0}.{5}.{5}}}\nonumber\\&+{\frac { \left( \Del{1}.{5}.{2} \right) ^{3} \left( \Del
{0}.{5}.{1} \right) ^{3}}{\Dell{0}.{5}.{1}.{5}\Del{1}.{5}.{1}\Del{1}.{
5}.{5}s_{{2,5}}\spa{2}.{3}\spa{3}.{4}\spa{4}.{5}s_{{1,5}}}}\nonumber\\&+{\frac {
 \left( \spa{1}.{2} \right) ^{3} \left( \Dell{0}.{3}.{0}.{2} \right) ^
{3}}{\sqrt {z_{{5}}}\spa{4}.{5}s_{{1,3}} \left( z_{{1}}+z_{{2}}+z_{{3}
} \right) \Del{0}.{3}.{3}\Del{0}.{3}.{4}s_{{1,2}}\Del{0}.{2}.{1}\Del{0
}.{2}.{2}\Del{0}.{2}.{3}}}\nonumber\\&+{\frac { \left( \spa{1}.{2} \right) ^{3}
 \left( \Dell{0}.{4}.{0}.{2} \right) ^{3}}{\sqrt {z_{{5}}}\spa{3}.{4}s
_{{1,4}} \left( 1 - z_5 \right) \Del{0}.{4}.{4
}\Del{0}.{4}.{5}s_{{1,2}}\Del{0}.{2}.{1}\Del{0}.{2}.{2}\Del{0}.{2}.{3}
}}\nonumber\\&+{\frac { \left( \spa{1}.{2} \right) ^{3} \left( \Dell{0}.{4}.{0}.{3
} \right) ^{3}}{\sqrt {z_{{5}}}\spa{2}.{3}s_{{1,4}} \left(
1-z_5  \right) \Del{0}.{4}.{4}\Del{0}.{4}.{5}s_{{1,3}}
\Del{0}.{3}.{1}\Del{0}.{3}.{3}\Del{0}.{3}.{4}}}\nonumber\\&-{\frac { \left( \spa{1
}.{2} \right) ^{3} \left( \Dell{0}.{5}.{0}.{2} \right) ^{3}}{\spa{3}.{
4}\spa{4}.{5}s_{{1,5}}\Del{0}.{5}.{5}s_{{1,2}}\Del{0}.{2}.{1}\Del{0}.{
2}.{2}\Del{0}.{2}.{3}}}\nonumber\\&-{\frac { \left( \Dell{0}.{5}.{0}.{3} \right) ^
{3} \left( \spa{1}.{2} \right) ^{3}}{\Del{0}.{3}.{4}s_{{1,3}}\Del{0}.{
3}.{1}\spa{2}.{3}\spa{4}.{5}s_{{1,5}}\Del{0}.{5}.{5}\Del{0}.{3}.{3}}}\nonumber\\&-
{\frac { \left( \Dell{0}.{5}.{0}.{4} \right) ^{3} \left( \spa{1}.{2}
 \right) ^{3}}{\Del{0}.{4}.{5}\Del{0}.{4}.{1}\spa{2}.{3}\spa{3}.{4}s_{
{1,5}}\Del{0}.{5}.{5}s_{{1,4}}\Del{0}.{4}.{4}}}.
\end{align}

\subsubsection{$n=6$}
\label{sec:4.2.6}

Finally, for six collinear gluons there are $2^7 = 128$ different splitting
amplitudes, which can be expressed by 23 independent ones. To find all
independent amplitudes we have to use Eq.~(\ref{mmmnpm}) for the first
time. Due to the length of the results we give here only two examples
obtained with the help of Eqs.~(\ref{mnpp}) and (\ref{mmnpm}),
\begin{align}
&\ssplit{1^-, 2^+,  3^+, 4^+, 5^+, 6^+}{+}=\nonumber\\&
{\frac { \left( \Del{0}.{2}.{1} \right) ^{3}\sqrt {z_{{1}}}}{\sqrt {z_
{{1}}z_{{6}}}\spa{1}.{2}\spa{3}.{4}\spa{4}.{5}\spa{5}.{6}s_{{1,2}}
 \left( z_{{1}}+z_{{2}} \right) \Del{0}.{2}.{2}\Del{0}.{2}.{3}}}\nonumber\\&+{
\frac { \left( \Del{0}.{3}.{1} \right) ^{3}\sqrt {z_{{1}}}}{\sqrt {z_{
{1}}z_{{6}}}\spa{1}.{2}\spa{2}.{3}\spa{4}.{5}\spa{5}.{6}s_{{1,3}}
 \left( z_{{1}}+z_{{2}}+z_{{3}} \right) \Del{0}.{3}.{3}\Del{0}.{3}.{4}
}}\nonumber\\&+{\frac { \left( \Del{0}.{4}.{1} \right) ^{3}\sqrt {z_{{1}}}}{\sqrt
{z_{{1}}z_{{6}}}\spa{1}.{2}\spa{2}.{3}\spa{3}.{4}\spa{5}.{6}s_{{1,4}}
 \left( z_{{1}}+z_{{2}}+z_{{3}}+z_{{4}} \right) \Del{0}.{4}.{4}\Del{0}
.{4}.{5}}}\nonumber\\&+{\frac { \left( \Del{0}.{5}.{1} \right) ^{3}\sqrt {z_{{1}}}
}{\sqrt {z_{{1}}z_{{6}}}\spa{1}.{2}\spa{2}.{3}\spa{3}.{4}\spa{4}.{5}s_
{{1,5}} \left( z_{{1}}+z_{{2}}+z_{{3}}+z_{{4}}+z_{{5}} \right) \Del{0}
.{5}.{5}\Del{0}.{5}.{6}}}\nonumber\\&-{\frac { \left( \Del{0}.{6}.{1} \right) ^{3}
}{\spa{1}.{2}\spa{2}.{3}\spa{3}.{4}\spa{4}.{5}\spa{5}.{6}s_{{1,6}}\Del
{0}.{6}.{6}}},\\\nonumber\\
&\ssplit{1^-, 2^-,  3^+, 4^+, 5^+, 6^+}{-}=\nonumber\\&
{\frac {{z_{{1}}}^{2} \left( \Del{1}.{3}.{2} \right) ^{3}}{\sqrt {z_{{
1}}z_{{6}}}\spa{2}.{3}\spa{4}.{5}\spa{5}.{6}s_{{2,3}}\Del{1}.{3}.{1}
\Del{1}.{3}.{3}\Del{1}.{3}.{4}}}\nonumber\\&+{\frac {{z_{{1}}}^{2} \left( \Del{1}.
{4}.{2} \right) ^{3}}{\sqrt {z_{{1}}z_{{6}}}\spa{2}.{3}\spa{3}.{4}\spa
{5}.{6}s_{{2,4}}\Del{1}.{4}.{1}\Del{1}.{4}.{4}\Del{1}.{4}.{5}}}\nonumber\\&+{
\frac {{z_{{1}}}^{2} \left( \Del{1}.{5}.{2} \right) ^{3}}{\sqrt {z_{{1
}}z_{{6}}}\spa{2}.{3}\spa{3}.{4}\spa{4}.{5}s_{{2,5}}\Del{1}.{5}.{1}
\Del{1}.{5}.{5}\Del{1}.{5}.{6}}}\nonumber\\&-{\frac {{z_{{1}}}^{2} \left( \Del{1}.
{6}.{2} \right) ^{3}\sqrt {z_{{6}}}}{\sqrt {z_{{1}}z_{{6}}}\spa{2}.{3}
\spa{3}.{4}\spa{4}.{5}\spa{5}.{6}s_{{2,6}}\Del{1}.{6}.{1}\Del{1}.{6}.{
6} \left( z_{{2}}+z_{{3}}+z_{{4}}+z_{{5}}+z_{{6}} \right) }}\nonumber\\&+{\frac {
 \left( \spa{1}.{2} \right) ^{3} \left( z_{{1}}+z_{{2}} \right) ^{3}
\sqrt {z_{{1}}}}{\sqrt {z_{{1}}z_{{6}}}\spa{3}.{4}\spa{4}.{5}\spa{5}.{
6}s_{{1,2}}\Del{0}.{2}.{1}\Del{0}.{2}.{2}\Del{0}.{2}.{3}}}\nonumber\\&+{\frac {
 \left( \spa{1}.{2} \right) ^{3} \left( z_{{1}}+z_{{2}}+z_{{3}}
 \right) ^{3}\sqrt {z_{{1}}}}{\sqrt {z_{{1}}z_{{6}}}\spa{2}.{3}\spa{4}
.{5}\spa{5}.{6}s_{{1,3}}\Del{0}.{3}.{1}\Del{0}.{3}.{3}\Del{0}.{3}.{4}}
}\nonumber\\&+{\frac { \left( \spa{1}.{2} \right) ^{3} \left( z_{{1}}+z_{{2}}+z_{{
3}}+z_{{4}} \right) ^{3}\sqrt {z_{{1}}}}{\sqrt {z_{{1}}z_{{6}}}\spa{2}
.{3}\spa{3}.{4}\spa{5}.{6}s_{{1,4}}\Del{0}.{4}.{1}\Del{0}.{4}.{4}\Del{0
}.{4}.{5}}}\nonumber\\&+{\frac { \left( \spa{1}.{2} \right) ^{3} \left( z_{{1}}+z_
{{2}}+z_{{3}}+z_{{4}}+z_{{5}} \right) ^{3}\sqrt {z_{{1}}}}{\sqrt {z_{{
1}}z_{{6}}}\spa{2}.{3}\spa{3}.{4}\spa{4}.{5}s_{{1,5}}\Del{0}.{5}.{1}
\Del{0}.{5}.{5}\Del{0}.{5}.{6}}}\nonumber\\&-{\frac { \left( \spa{1}.{2} \right) ^
{3}}{\spa{2}.{3}\spa{3}.{4}\spa{4}.{5}\spa{5}.{6}s_{{1,6}}\Del{0}.{6}.
{1}\Del{0}.{6}.{6}}}.
\end{align}
\label{sec:expl}

\section{Conclusion}
\label{sec:concl}

In this paper we have considered the collinear limit of multi-gluon QCD amplitudes at tree level. We
have used the new MHV rules for constructing colour ordered amplitudes from MHV vertices 
together with the general collinear
  factorization
  formula to derive timelike splitting functions
  that are valid for  specific numbers of negative helicity gluons with an
arbitrary number of positive helicity gluons (or vice versa). In this limit, the full amplitude
factorises  into an MHV vertex multiplied by a multi-collinear splitting function that depends on the
helicities of the collinear gluons. These splitting functions are  
derived directly using MHV rules. Out of the full set of
MHV-diagrams contributing to the full amplitude, only the subset of MHV-diagrams which contain
an internal propagator which goes on-shell in the multi-collinear limit contribute.

We find that the splitting functions can be characterised by $\Delta M$, the  difference between the
number of negative helicity gluons before taking the collinear limit, and the number after.   $\Delta
M+1$ also coincides with the number of MHV vertices involved in the splitting functions.   Our main
results are splitting functions for arbitrary numbers of gluons where $\Delta M  =0,1,2$. Splitting
functions where the difference in the number of positive helicity gluons $\Delta P = 0,1,2$ are
obtained by the parity transformation.   These general results are sufficient to describe {\em all} collinear
limits with up to six gluons.   We have given explicit results for up to four collinear gluons for all
independent helicity combinations, which numerically agree with the results of Ref.~\cite{delduca},
together with new results for five and six collinear gluons. This method could be applied to
higher numbers of negative helicity gluons, and via the MHV-rules for quark vertices, to the collinear
limits of quarks and gluons.

We anticipate that the results presented here will be useful in developing higher order perturbative
predictions for observable quantities, such as jet cross sections at the LHC.

\section{Acknowledgements}
EWNG and VVK acknowledge the support of PPARC through
Senior Fellowships and TGB acknowledges the award of a PPARC studentship.

\end{document}